\documentclass[aps,prl,citeautoscript,reprint,superscriptaddress,floatfix,footinbib,showkeys]{revtex4-2}
\usepackage{setspace}
\usepackage[utf8]{inputenc}
\usepackage[colorlinks=true,linkcolor=black,urlcolor=black,filecolor=black,citecolor=black]{hyperref}
\usepackage{color,soul}
\usepackage{amsmath}
\usepackage{amsfonts}
\usepackage{amssymb}
\usepackage{graphicx}
\usepackage{dcolumn}
\usepackage{bm}
\usepackage{subfigure}
\usepackage{booktabs}

\newcommand{\mr}[1]{\mathrm{#1}}

\begin{document}
\title{Bolometric detection of Josephson \textcolor{black}{inductance} in a highly \textcolor{black}{resistive} environment}
\author{Diego Subero}\email{diego.suberorengel@aalto.fi}
\affiliation{PICO Group, QTF Centre of Excellence, Department of Applied Physics, Aalto University School of Science, P.O. Box 13500, 0076 Aalto, Finland}
\author{Olivier Maillet}
\affiliation{PICO Group, QTF Centre of Excellence, Department of Applied Physics, Aalto University School of Science, P.O. Box 13500, 0076 Aalto, Finland}
\affiliation{Université Paris-Saclay, CEA, CNRS, SPEC, 91191 Gif-sur-Yvette, France}

\author{Dmitry S. Golubev}
\author{George Thomas}
\author{Joonas T. Peltonen}

\affiliation{PICO Group, QTF Centre of Excellence, Department of Applied Physics, Aalto University School of Science, P.O. Box 13500, 0076 Aalto, Finland}
\author{Bayan Karimi}
\affiliation{PICO Group, QTF Centre of Excellence, Department of Applied Physics, Aalto University School of Science, P.O. Box 13500, 0076 Aalto, Finland}
\affiliation{QTF Centre of Excellence, Department of Physics, Faculty of Science, University of Helsinki, 00014 Helsinki, Finland}
\author{Marco Marín-Suárez}

\affiliation{PICO Group, QTF Centre of Excellence, Department of Applied Physics, Aalto University School of Science, P.O. Box 13500, 0076 Aalto, Finland}

\author{Alfredo Levy Yeyati}

\author{Rafael Sánchez}

\author{Sunghun Park}

\affiliation{Departamento de Física Teórica de la Materia Condensada, Condensed Matter Physics Center (IFIMAC) and Instituto Nicolás Cabrera, Universidad Autonoma de Madrid, 28049 Madrid, Spain}

\author{Jukka P. Pekola}
\affiliation{PICO Group, QTF Centre of Excellence, Department of Applied Physics, Aalto University School of Science, P.O. Box 13500, 0076 Aalto, Finland}


\maketitle

\section{Abstract}

The Josephson junction is a building block of quantum circuits. Its behavior, well understood when treated as an isolated entity, is strongly affected by coupling to an electromagnetic environment.
In 1983, Schmid predicted that a Josephson junction shunted by a resistance exceeding the resistance quantum  $ \textit{R}_\mr{Q} = \textit{h}/4\textit{e}^2 \approx 6.45$ k$\Omega$ for Cooper pairs would become insulating since the phase fluctuations would destroy the coherent Josephson coupling.  However, recent microwave measurements have questioned this interpretation. Here, we insert a small Josephson junction in a Johnson-Nyquist-type setup where it is driven by weak current noise arising from thermal fluctuations. Our heat probe minimally perturbs the junction's equilibrium, shedding light on features not visible in charge transport. We find that the Josephson critical current completely vanishes in DC charge transport measurement, and the junction demonstrates Coulomb blockade in agreement with the theory. Surprisingly, thermal transport measurements show that the Josephson junction acts as an inductor at high frequencies, unambiguously demonstrating that a supercurrent survives despite the Coulomb blockade observed in DC measurements.

\section{Introduction}

\begin{figure*}[ht!]
	\centering
	\includegraphics[width=\textwidth]{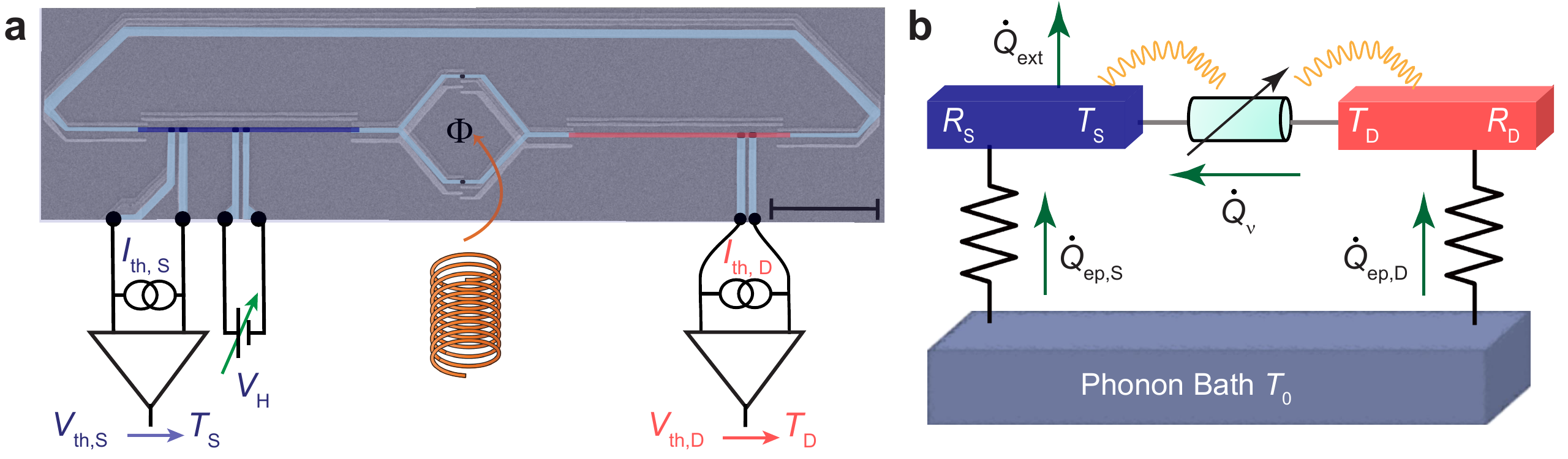}
	\caption{\textbf{Experimental setup and principle of the photonic heat transport in high ohmic environment}. \textbf{a} Colored scanning electron micrograph (scale bar: 5 $\mu$m) highlighting the \textcolor{black}{(Cr)} normal metal (blue and red) and the aluminum superconducting leads (light blue). The Josephson energy of the SQUID is tuned with an external magnetic field. Aluminum leads (vertical, light blue) are connected through an oxide tunnel barrier to the Cr-strip to cool down its electrons locally or as an electronic temperature sensor using a floating DC current source. \textbf{b} Schematic illustration of the thermal model of the system. The drain-source heat flow $\dot{Q}_{\nu}$ is adjusted by the SQUID. The source and drain electron baths are thermally coupled to the phonon bath (which here is hotter), receiving a power $\dot{Q}_\mr{ep, S}$ and $\dot{Q}_\mr{ep, D}$, respectively. Wiggly lines highlight the \textcolor{black}{strong} interaction between the SQUID and the Ohmic environment, mediated by photons. }\label{Fig1}
\end{figure*}

Thermal transport by photons in electrical circuits arises from Johnson-Nyquist noise \cite{nyquist1928thermal,johnson1928thermal} between two resistive elements at unequal temperatures. The resulting current noise flowing in the ideally lossless circuit linking the two resistors provides an efficient thermalization and energy exchange \textcolor{black}{channel} at very low temperatures \cite{Schmidt2004, MatthiasMeschkeandPekolaJukka2006}. This current noise can be modulated by adding a suitable tunable dissipationless element in the circuit, which can conveniently be implemented by a magnetically \cite{MatthiasMeschkeandPekolaJukka2006,partanen2018flux,Ronzani2018} or electrically \cite{maillet2020electric} controlled Josephson device. So far, both theoretical \cite{Ojanen2008, Pascal2011} and experimental approaches \cite{MatthiasMeschkeandPekolaJukka2006, partanen2018flux, Ronzani2018, maillet2020electric} of photonic heat transport have only considered on-chip resistors with resistance $R$ much smaller than the superconducting resistance quantum $R_\mr{Q}$, where the environmental back-action effect on the junctions is weak.
More fundamentally, energy transport \textcolor{black}{at different frequencies} through quantum coherent systems strongly coupled to a \textcolor{black}{resistive} environment, 
\textcolor{black}{characterized by the dissipation strength $R_\mr{Q}/R$, remains largely unexplored \cite{thomas2019photonic,leger2019observation}}.

It is well-established that the electrical transport properties of a superconducting junction depend on the electromagnetic environment in which it is embedded. Charge transport through a tunnel junction in an Ohmic environment with resistance comparable to the resistance quantum is suppressed at low voltage bias and temperature because of Coulomb blockade \cite{ingold1992charge}, and the extension of this phenomenon to superconducting junctions, which are intrinsically phase-coherent, comes naturally \cite{kuzmin1991coulomb,averin1990incoherent,ingold1992charge,corlevi2006phase}.
Recent experiments using this effect include the production of antibunched photons at high rates \cite{grimm2019bright} and suppressing the Andreev bound state-induced zero bias anomaly, which could be beneficial in the arduous search for Majorana quasiparticles \cite{zhang2022suppressing}. For a Josephson junction shunted by a resistor with resistance exceeding $R_\mr{Q}$, the supercurrent peak (i.e., current at zero applied voltage bias) is predicted to disappear, being shifted to finite voltage as a result of inelastic Cooper pair tunneling \cite{ingold1992charge}. This is accompanied by a sub-linear current-voltage characteristic  \textcolor{black}{ as $I\sim V^{\frac{2R}{R_\mr{Q}}-1}$ at low bias voltages and extremely low temperature, specifically when $k_\mr{B}T\ll eV\ll R_\mr{Q}E_\mr{c}/R\pi$, with $k_\mr{B}$ the Boltzmann constant, $E_\mr{c}$ the charging energy, and $e$ the elementary charge. According to the theory, a phase transition should occur at $R = R_\mr{Q}$.} This transition, which can be associated with the one predicted by Schmid \cite{schmid1983diffusion} and Bulgadaev \cite{bulgadaev1984phase}, was first \textcolor{black}{tested} in DC charge transport experiments \cite{kuzmin1991coulomb,penttila1999superconductor,corlevi2006phase, yagi1997phase,penttila2001experiments}. However, recent admittance measurements of small junctions, shunted by a highly Ohmic environment \cite{murani2020absence} called into question the scenario of a dissipative phase transition, leading to further debate about the very existence of this transition \cite{Perti2020, JoyezReply, Masuki2022, Sepulcre2022, MasukiReply2022, kuzmin2023observation}.

In this context, we present a heat transport experiment in which a small tunable junction (effectively a superconducting quantum interference device SQUID) is embedded in a Johnson-Nyquist setup with hot and cold \textcolor{black}{resistors with} resistances $R >R_\mr{Q}$ to explore this regime. The SQUID geometry enables magnetic-flux control of the photonic heat current \cite{MatthiasMeschkeandPekolaJukka2006}. It is intended to demonstrate the destruction or resilience of the Josephson coupling through the observations of heat flow oscillations, or lack thereof if the junctions are truly insulating. We find that the magnitude of heat current flowing from one resistor to another remains close to the value given by the quantum limit, and it exhibits clear oscillations with the external magnetic flux, similar to the systems embedded in a low impedance environment \cite{MatthiasMeschkeandPekolaJukka2006}. While this observation might point towards the survival of \textcolor{black}{a supercurrent at high frequencies, a control experiment on DC charge transport (angular frequency $\omega = 0$) shows clear suppression of the charge current at low voltage bias caused by the environmental Coulomb blockade, in line with theoretical predictions and previous experimental results \cite{kuzmin1991coulomb,penttila1999superconductor}. } This apparent contradiction, highlighting the role of heat transport as a complementary probe when many-body correlations are present \cite{giazotto2012josephson, Sivre2018}, is discussed within the existing theoretical and experimental literature.

\begin{figure*}[ht!]
	\centering
	\includegraphics[width=1\textwidth]{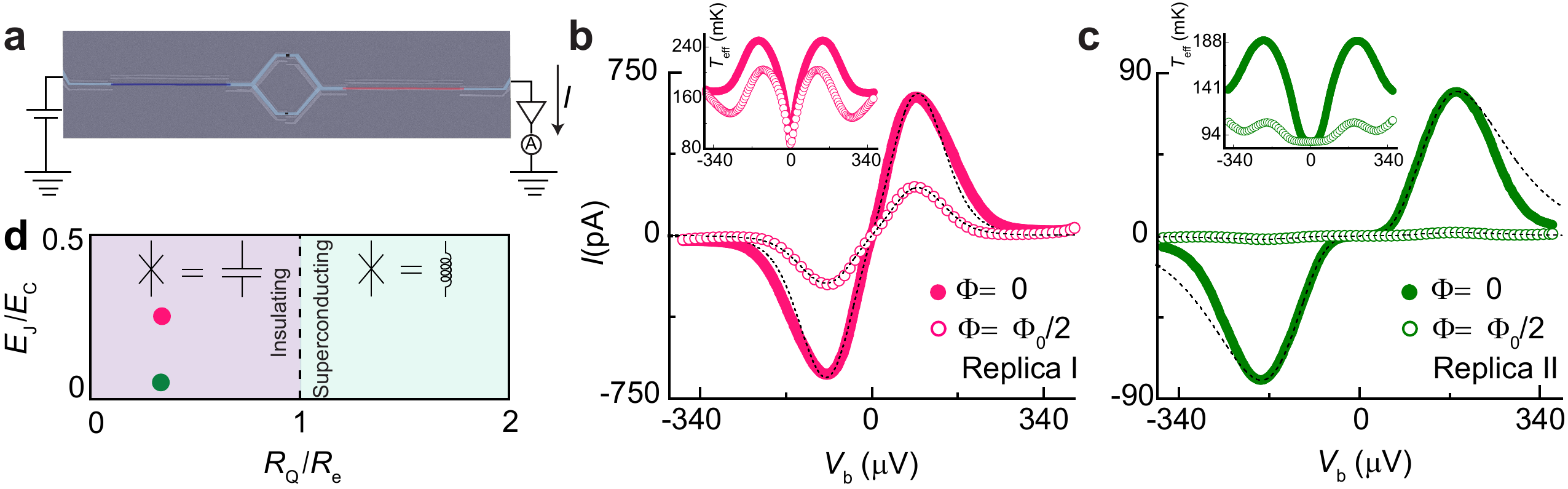}
	\caption{\textbf{DC charge transport measurements}. \textbf{a} Scanning electron micrograph of one of the Replica samples (scale bar: 5 $\mu$m) with the schematics of the IV measurement. \textbf{b, c} A close-up of the IVC at the low-bias voltage for \textcolor{black}{the two Replica samples} measured at a cryostat temperature of 87 mK exhibiting the Coulomb blockade feature. The measurements were recorded at two magnetic flux values $\Phi= 0$ (solid circles) and $\Phi= \Phi_0/2$ (\textcolor{black}{empty circles}). \textcolor{black}{The dashed lines in panels b and c are the theoretical results obtained by the standard \textit{P(E)}-theory for two different magnetic flux values. For Replica I in panel b, the fit parameters are: critical current $I_\mr{C}\sim 7$ nA, Josephson energy $E_\mr{J}\sim 0.17$ K, charging energy $E_\mr{C}\sim 0.6$ K and, for Replica II in panel \textbf{c}: $I_\mr{C}\sim 3$ nA, $E_\mr{J}\sim 0.08$ K, $E_\mr{C}= 1.4$ K.} \textcolor{black}{Cr-strips' resistance} is $R_\mr{e}= 11$ k$\Omega$ for both samples. \textcolor{black}{The inset in panels b and c show the effective temperature of the resistor at low voltage bias.} \textbf{d} Illustration of the Schmid phase diagram for a Josephson junction attached to a resistive environment $R$ at zero temperature. Here, $R= R_\mr{S} + R_\mr{D}= 2R_\mr{e}$ is the total resistance of the environment. Our samples are well placed in the insulating part, represented by the two points. }\label{Replica}
\end{figure*}

\section{Results}

\subsection{Experimental setup}

Our device (see Fig. \ref{Fig1}a  for an SEM image) consists of a SQUID between two nominally identical on-chip thin chromium \textcolor{black}{(Cr)} resistors acting as thermal baths, from now on referred to \textcolor{black}{as (hot) drain and (cold) source} with resistances denoted by $R_\mr{D}$ and $R_\mr{S}$, respectively. Each arm of the SQUID is galvanically connected to one source and drain resistor of volume $\Omega= 10\times0.1\times0.014$ $\mu$m$^3$ and whose resistance is nominally equal to that of an independently measured resistor with same dimensions on the same chip, with a value $R_\mr{S}=R_\mr{D}= 11\pm0.5$ k$\Omega$ (see Supplementary S1). The distance between the SQUID and the resistors is kept short (a few microns) to avoid suppression of environment-induced effects via stray capacitance. The series configuration of the SQUID and resistors is further closed into a loop by a superconducting line. This warrants efficient electromagnetic heat transport through improved impedance matching \cite{Timofeev2009}. The clean contact between chromium and superconducting aluminum leads \textcolor{black}{(see Supplementary S2)} serves as an Andreev-mirror \cite{andreev1965thermal}, which enables essentially perfect conversion to charge transport by Cooper pairs in the superconducting strips while effectively suppressing quasiparticle heat diffusion along them at low temperatures \textcolor{black}{($T\lesssim 0.2 T_\mr{c}\sim 260$ mK for aluminum)} \cite{Timofeev2009}. Four external superconducting leads are contacted with the source resistor through a thin oxide barrier, forming NIS-tunnel junctions. A pair of these junctions is used to measure the electronic temperature (in the case of quasi-equilibrium where the electron temperature is well-defined \cite{Giazotto2006}) by applying a small DC-current bias through it, whereas another pair is used to locally cool the resistor when voltage-biased \cite{Giazotto2006}. The electron temperature of the drain resistor is measured simultaneously by another SINIS junction structure, as depicted in Fig. \ref{Fig1}a. We have presented data on two samples, henceforth called Sample I and Sample II.

\subsection{DC measurements of the replica sample}

We first measure the current-voltage characteristics (IVC) of a reference sample (see Fig. \ref{Replica}a) on the same chip\textcolor{black}{, made during the same fabrication run,} from now on called ``Replica", with nominally equal parameters as the main sample. This provides estimates of parameters for the heat transport experiments and enables comparison between charge and heat transport behavior. \textcolor{black}{Note that although the geometries of the samples used for these measurements are slightly different, the central part of the two samples (resistors + SQUID) are nominally identical. In the sample used for the heat transport experiment, the superconducting loop keeps the current noise flowing within a closed loop and helps us define the noise power transmission coefficient.}

Figure \ref{Replica}b and \ref{Replica}c show the IVC for the two Replica samples at a phonon temperature of $T_\mr{0}= 87$ mK in the low bias region at two different magnetic flux values $\Phi= 0$ (solid circles) and $\Phi=  \Phi_\mr{0}/2$ (open circles) with $\Phi_0 = \pi\hbar/e$ the superconducting magnetic flux quantum. Suppression with respect to the unblocked case is observed in the low-bias DC current through the SQUID, which is more robust for Replica II (panel c) due to its higher charging energy $E_\mr{C}$. This observation is well understood in the framework of dynamical Coulomb blockade \cite{ingold1992charge}: the resistive environment impedes charge relaxation after a Cooper-pair tunneling event through junctions with high charging energy, which translates to a conductance reduction at low energy. The \textcolor{black}{value} of $E_\mr{C}$ (0.6 K for Replica I and 1.4 K for Replica II) is extracted from the current peak feature appearing in the IVC, which for small Josephson junctions in contact with an \textcolor{black}{resistive} environment with \textcolor{black}{$R > R_\mr{Q}$} occurs at a voltage bias $eV_\mr{b}\sim 2E_\mr{C}$ \cite{devoret1990effect,kuzmin1991coulomb}. 
The Josephson energy is estimated by using the Ambegaokar-Baratoff relation as \textcolor{black}{ $E^{AB}_\mr{J} = \Phi_0\Delta/4eR_\mr{J}$} (0.07 K for Replica I and 0.03 K for Replica II), with $\Delta\approx 200$ $\mu e$V the aluminum superconducting energy gap and $R_\mr{J}$ the quasiparticle tunnel resistance of the SQUID. \textcolor{black}{This resistance is obtained experimentally, see Supplementary S3 for more details}. 

The dashed lines in Fig. \ref{Replica}b and \ref{Replica}c are the theoretical results obtained by the standard \textit{P(E)} theory \cite{ingold1992charge} \textcolor{black}{under the condition $E_\mr{J}\ll k_\mr{B}T$} in an \textit{RC}-environment, which highlights the effect of the electromagnetic environment on Josephson phase fluctuations (see Supplementary S4).
\textcolor{black}{Despite the fact that the condition $E_\mr{J}\ll k_\mr{B}T$ is not well satisfied for either of the two samples, one can use a simple rule: if the Josephson energy $E_\mr{J}$ is less than Josephson energy obtained from Ambegaokar-Baratoff $E^{AB}_\mr J$ (which is our case), then $P(E)$ theory should still hold regardless of the condition $E_\mr{J}\ll k_\mr{B}T$ and the relatives' magnitudes of $E_\mr{J}$ and $E_\mr{C}$ \cite{lu2023phase}.}
\textcolor{black}{In these calculations, we include overheating caused by the applied bias, as depicted in the inset of Figs. \ref{Replica}b and \ref{Replica}c, resulting in good agreement with the data at low voltage bias. However, a noticeable discrepancy is observed at high voltages for Replica II. A possible reason for this discrepancy could be that, at voltage beyond $V_\mr{b}\sim 2E_\mr{C}/e$, the Josephson frequency is high, $eV/\pi\hbar>100$ GHz: in this range, the impedance of the environment may significantly differ from our simple RC model. Furthermore, at these voltages closer to the superconducting gap $V\sim 2\Delta/e$ quasiparticles can contribute but are not included in \textit{P(E)} theory. On the other hand, the data at $\Phi= \Phi_0/2$ is fitted by considering an asymmetry factor of the SQUID critical current {$d = 0.58$} for Replica I and $d = 0.15$ for Replica II, respectively. Note that the Josephson energy $E_\mr{J}$ is a fitting parameter in the model.}
Finally, Fig. \ref{Replica}d shows the Schmid phase diagram with the two points related to each Replica sample, showing their insulating character as expected.

\subsection{Heat transport measurements}

With these results as a reference, we now turn to heat transport measurements, the main focus of this work. The power flowing from drain to source $\dot{Q}_\nu$ under a thermal gradient is determined by measuring the electronic temperatures of the drain $T_\mr{D}$ and the source $T_\mr{S}$. The temperature difference is generated by DC biasing the source resistor with a voltage $V_\mr{H} \lesssim 2\Delta/e$ that enables electronic cooling of the source by removal of hot electrons \cite{Giazotto2006, Timofeev2009}. \textcolor{black}{This effect can be observed in Fig. 3a and 3b, for the two samples at a fixed temperature $T_\mr{0}$} in each case. In steady-state, these temperatures involve the different energy relaxation channels in the system, as illustrated in the thermal model that accounts for our setup shown in Fig. \ref{Fig1}b. By \textcolor{black}{energy conservation}, a direct relation (\textcolor{black}{valid at temperatures $T/T_\mr{C}< 0.2$ when quasiparticle heat diffusion along the superconductor is exponentially suppressed \cite{Timofeev2009})}, between $\dot{Q}_{\nu}$ and the temperatures ($T_\mr{S}, \ T_\mr{D} , \ T_\mr{0}$) measured in the system is found

\begin{equation}
  \dot{Q}_\nu (T_\mathrm{S}, T_\mathrm{D}, \Phi)= \dot{Q}_\mathrm{ep, D}(T_{\mr{D}}, T_\mr{0}), \label{Experimental equation}
\end{equation}

\noindent where $\dot{Q}_\mathrm{ep, D}= \Sigma\Omega[T_\mr{0}^\mr{5.93}- T_\mathrm{D}^\mr{5.93}] $ is the electron-phonon heat current governed by the drain resistor. Here, $\Sigma$ is the electron-phonon coupling constant of the normal metal, which was measured independently to be $\Sigma = (12\pm0.25)\times10^\mr{9}$ WK$^\mr{-5.93}$m$^\mr{-3}$, \textcolor{black}{see Supplementary S1}. With our experimental setup (see Fig. \ref{Fig1}a), we have full control of all temperatures and, therefore, the powers involved in the system, leading to an accurate and fully calibrated measurement of the thermal conductance between the drain and source.

\begin{figure}[ht!]
	\centering	\includegraphics[width=\columnwidth]{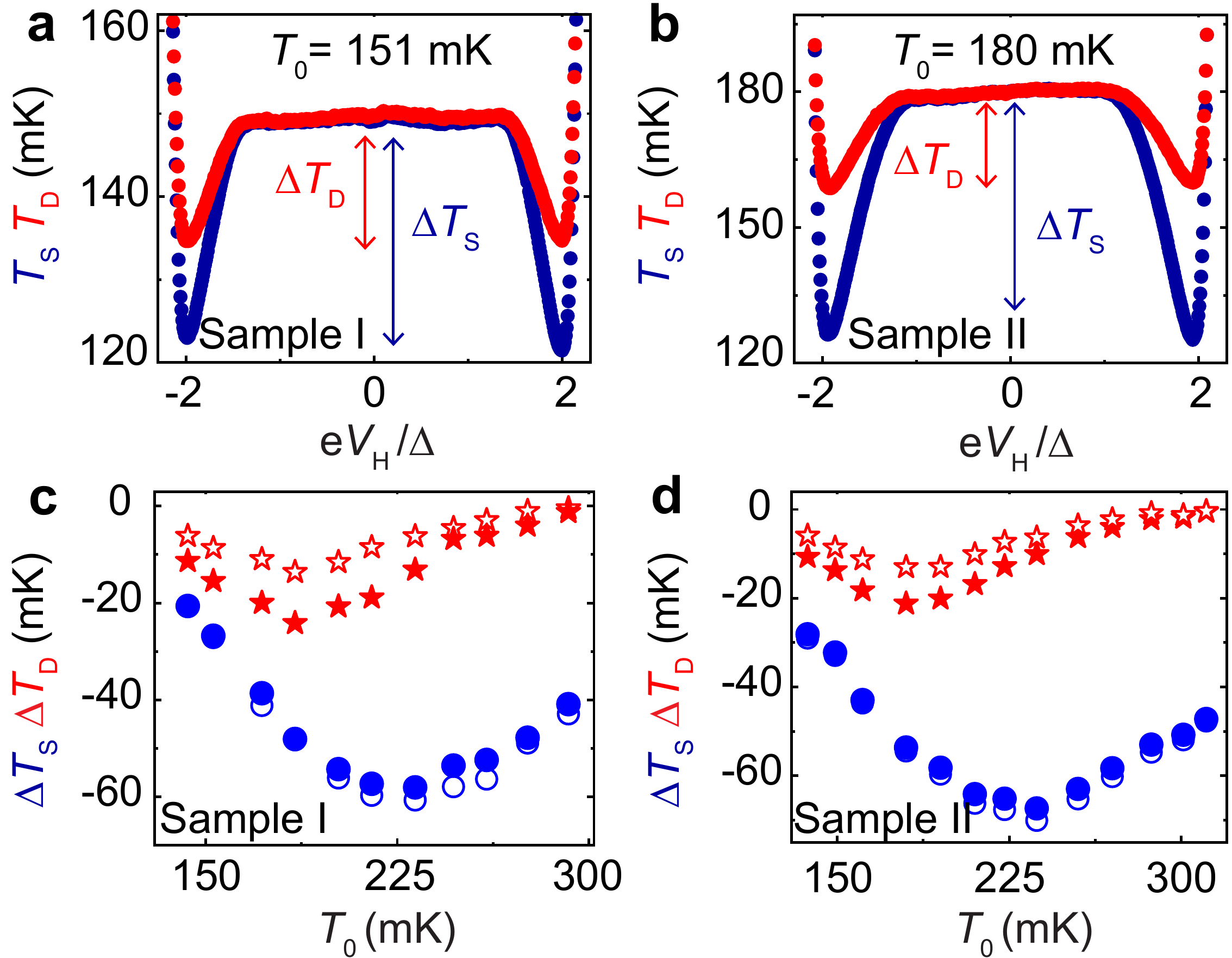}
	\caption{\textbf{Electronic refrigeration}. \textbf{a, b} \textcolor{black}{Electronic temperature of the source $T_\mr{S}$ (blue points) and drain $T_{\rm D}$ resistor (red points) at $\Phi= 0$ for samples I and II, respectively, as a function of the heating voltage $V_\mr{H}$ applied on the source resistor. The measurements were recorded at phonon temperature $T_\mr{0}=$ 151 mK (Sample I) and 180 mK (Sample II). \textbf{c, d}.- Temperature drops of the source $\Delta T_\mr{S}$ (circles) and the drain $\Delta T_\mr{D}$ (stars) recorded at two magnetic flux values $\Phi=0$ (filled symbols) and $\Phi=\Phi_0/2$ (empty symbols) against $T_\mr{0}$. The error bars are not shown since their values are smaller than the markers.}}\label{Figure3}
\end{figure}

\textcolor{black}{Figure \ref{Figure3}a and \ref{Figure3}b show the measured electronic temperature of the source and drain resistor with the applied voltage bias $V_\mr{H}$ for each sample at a representative phonon temperature $T_\mr{0}$. Notably, both samples show a significant decrease in the source and drain electronic temperature when the voltage approaches $V_\mr{H}^\mr{opt}\lesssim 2\Delta/e$, indicating that the cooling power of the SINIS refrigerator reaches its maximum \cite{Giazotto2006}.}
Figure \ref{Figure3}c and \ref{Figure3}d show the temperature drops $\Delta T_\mr{i} = T_\mr{i}(V_\mr{H}^\mr{opt}) - T_\mr{i}(V_\mr{H}= 0)$, i= S, D at the maximum cooling bias of the SINIS for the two samples at two \textcolor{black}{different} magnetic flux values.  At $V_\mr{H}= 0$, the electronic temperature equals the phonon temperature $T_\mr{0}$. These drops characterize the thermal coupling between the drain and source, i.e., the thermal conductance from drain to source. Source and drain temperature drops differ for the two magnetic fluxes supplied. \textcolor{black}{This difference is more evident in the temperature drops of the drain, where the drop at $\Phi = \Phi_0/2$ is noticeably weaker than at $\Phi = 0$ at low temperatures, indicating that the photonic channel effectively dominates the transport mechanism at lower temperatures \cite{Timofeev2009}.} The significant flux-tunability of the remote cooling process is a characteristic feature of SQUID interference \cite{MatthiasMeschkeandPekolaJukka2006}, which suggests that the environmental back-action \textcolor{black}{does not} destroy the Josephson coupling \textcolor{black}{at zero DC voltage bias}. As $T_\mr{0}$ increases above 200 mK, the photon thermal coupling gets smaller with respect to electron-phonon coupling.

To quantify the power transferred from the drain to the source resistor through a photon channel, the electronic temperatures measured on the two baths have been converted to heat current $\dot{Q}_\nu$ by using Eq. (\ref{Experimental equation}) and compared with the maximum power that can be transmitted through a single ballistic channel given by $\dot{Q}_\mr{Q} = \frac{\pi k_\mr{B}^{2}}{12\hbar}(T_\mr{D}^{2} - T_\mr{S}^{2})$ \cite{pendry1983quantum}. 

\begin{figure*}[ht!]
	\centering
	\includegraphics[width=1\textwidth]{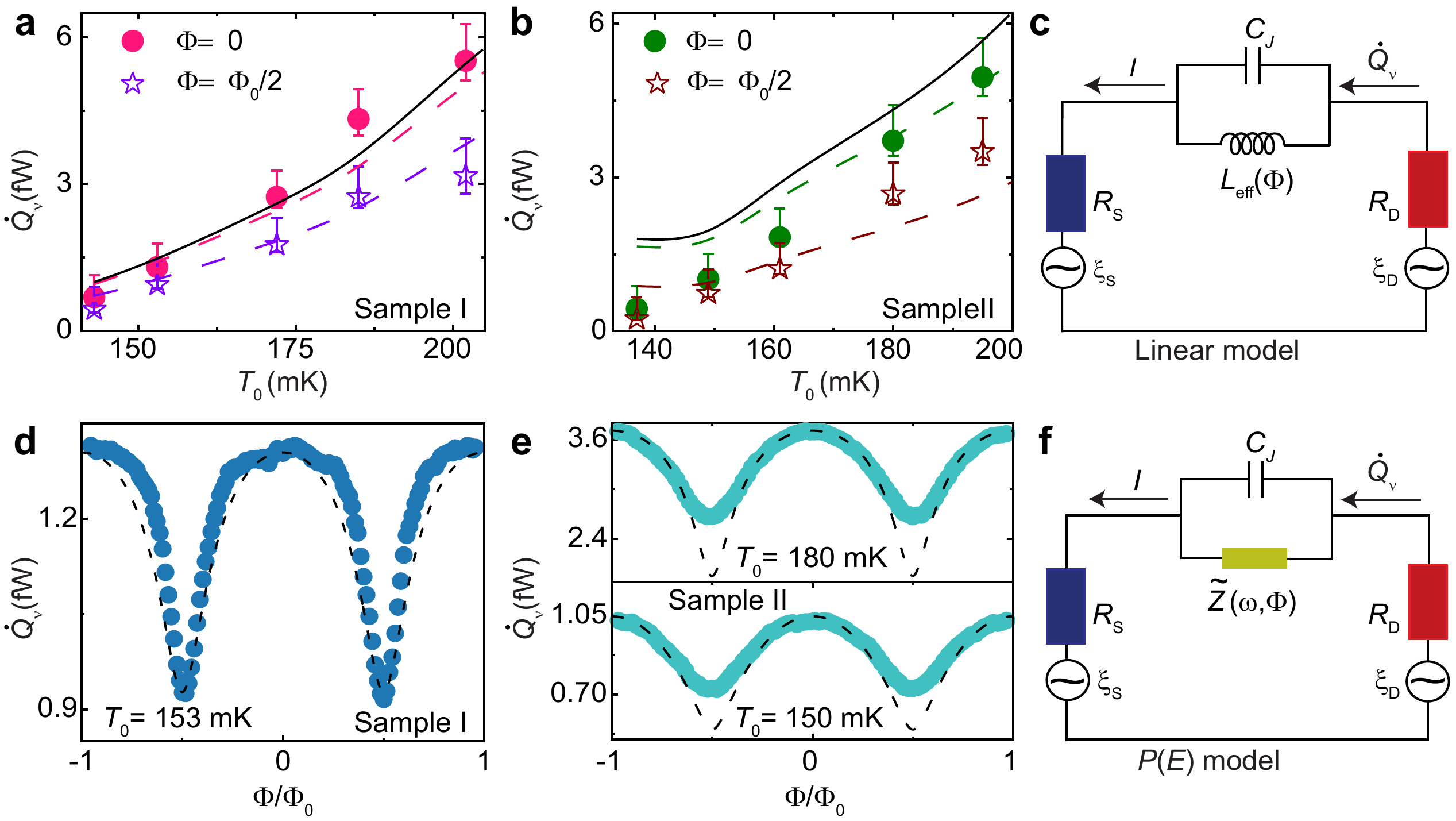}
	\caption{\textbf{Heat transport mediated by photons and the theoretical model proposed.} \textbf{a, b} Photonic heat current from the drain to the source by using the continuity equation (\ref{Experimental equation}). The error bars are given in their lower and upper parts, the combination of the thermometer calibration and electron-phonon coupling constant uncertainties, while the upper part also includes the parasitic heat leak on the resistors due to the NIS junctions, with a 0.4 fW upper bound estimate. The solid line represents the power transmitted through a single channel at the quantum limit $\dot{Q}_\mr{Q}$ (see text). The dashed lines are obtained by solving Eq. (\ref{Landauer}) \textcolor{black}{with a photon transmission probability calculated with the linear circuit model}, with the circuit parameters obtained from fitting the IVC of the Replica samples in \textcolor{black}{figure \ref{Replica}b and \ref{Replica}c}. \textbf{d, e} Heat current modulation as a function of the reduced magnetic flux $\Phi/\Phi_0$ through the SQUID at a given temperature $T_\mr{0}$. The dashed lines are the \textcolor{black}{application of the linear model (see text),} keeping the same circuit parameters as in \textcolor{black}{panels} a and b. \textbf{c} Electric representation of the device in the linear model where the SQUID is approximated by a re-normalized variable inductor $L_\mr{eff}(\Phi)$ in parallel with the junction geometric capacitor $C_\mr{J}$, and \textbf{f} \textit{P(E)} model with an effective impedance $\tilde{Z}(\omega,\Phi)$ replacing the Josephson element.}\label{Conductance}
\end{figure*}

The photonic heat current for the two samples in the temperature range of 140 - 200 mK is shown in Figs. \ref{Conductance}a and \ref{Conductance}b. For Sample I at $\Phi=0$ (solid circles), the energy is transferred from the drain to the source at a rate very close to that dictated by the quantum of thermal conductance (solid black line). Nevertheless, a minor deviation of $\dot{Q}_\nu$ from the quantum limit prediction is seen for Sample II. \textcolor{black}{As expected, as soon as the magnetic flux is switched on and reaches} the half flux quantum $\Phi = \Phi_0/2$ (\textcolor{black}{star symbols}), the heat current $\dot{Q}_\nu$ for both samples tends to decrease due to weaker photonic coupling \cite{Timofeev2009}.

We then confront the data with theoretical calculations based on the Landauer relation for heat current from drain to source \cite{Schmidt2004},

\begin{equation}
\dot{Q}_\nu = \int_{0}^{\infty} \frac{d\omega}{2\pi} \hbar\omega\tau(\omega,\Phi)\Big[\frac{1}{e^{\hbar\omega/k_\mr{B}T_\mr{D}}-1}-\frac{1}{e^{\hbar\omega/k_\mr{B}T_\mr{S}}-1}\Big] , \label{Landauer}
\end{equation}

\noindent where $\tau(\omega,\Phi)$ is the transmission probability of the thermal radiation from the drain to the source at angular frequency $\omega$. In the limit of small phase fluctuations around a given average phase bias $\varphi$, the SQUID can conveniently be approximated by a harmonic oscillator. In electrical terms, this translates to an effective phase-dependent Josephson inductance $L_\mr{eff}(\Phi) = \hbar/(2e|I_\mr{c}(\Phi)|\langle\cos\varphi\rangle)$ in parallel with a capacitance (see Fig. \ref{Conductance}c), where $\langle\cos\varphi\rangle$ is an average over phase fluctuations \cite{joyez2013self,Masuki2022}. Within this linear model, and by assuming the lumped approximation (valid since the dominant radiation wavelength $\lambda_\mr{th}=hc/k_\mr{B}T \sim 10$ cm at 150 mK is much larger than the circuit characteristic dimensions $\sim 50$ $\mu$m), the power transmission coefficient can be explicitly written \cite{Schmidt2004,Pascal2011} as $\tau(\omega,\Phi) = 4R_\mr{S}R_\mr{D}/|Z_\mr{T}(\omega,\Phi)|^2$ (see methods), with $Z_\mr{T}(\omega, \Phi)$ the frequency-dependent total series impedance of the circuit. In this framework, the maximum heat transfer is expected for a perfect impedance matching when $R_\mr{S}$ =  $R_\mr{D}$ and when the phase fluctuations are small, i.e., $\langle\cos\varphi\rangle\simeq 1$ under no net electrical bias. 
However, for strong phase fluctuations at high resistances $R_\mr{S}+R_\mr{D} >R_\mr{Q}$, one naively expects the average value of the cosine almost to vanish, $\langle\cos\varphi\rangle\approx 0$. Indeed, we have shown above that the \textcolor{black}{DC} charge transport measurements of the Replica are well described by the usual $P(E)$ theory of Coulomb blockade, which implies $\langle\cos\varphi\rangle \approx 0$. Assuming this, the SQUID can be regarded as a parallel connection of a capacitor $C_J$ and of high effective impedance  $\tilde{Z}(\omega,\Phi)\propto 1/I_\mr{c}^2(\Phi)$, see Fig. \ref{Conductance}f. Then the modulation of the Josephson coupling by flux almost does not affect the transmission probability $\tau(\omega,\Phi)$, and the oscillations of the heat flow are expected to be very small. Below we will show that the strong modulation of the heat flux observed in our experiment is consistent with the assumption of nonvanishing $\langle\cos\varphi\rangle$ rather than with $\langle\cos\varphi\rangle\approx 0$. \textcolor{black}{This finding is similar to that in the recent study conducted by Pechenezhskiy \textit{et. al.,} \cite{pechenezhskiy2020superconducting} on a fluxonium qubit shunted by a high impedance environment in which they have revealed the presence of non-vanishing ground state renormalization $\langle 0 |\cos{\varphi}| 0 \rangle$ in the divergence inductance limit and the Bloch band structure of a single Josephson junction with $E_\mr{J}\sim E_\mr{C}$, which leads to modulations in the transition frequency with the external magnetic flux. Our work, combined with theirs, will be useful in developing a general theory that can clarify  the dynamics of a Josephson junction at finite frequencies in a high impedance environment.}

The dashed lines displayed in figures \ref{Conductance}a and \ref{Conductance}b are the theoretical results \textcolor{black}{obtained by solving Eq. (\ref{Landauer})} within the linear model for the corresponding magnetic fluxes applied. For Sample I, reasonable agreement with the experimental data at $\Phi= 0$ is found if we use the bare Josephson junction inductance \textcolor{black}{$L(\Phi) = \frac{\hbar}{2eI_\mr{c}(\Phi)}$} in the calculated $Z_\mr{T}(\omega, \Phi)$. Nevertheless, for Sample II, a significant deviation from the data is observed as the temperature is lowered. This deviation can be captured if we set the re-normalization parameter \textcolor{black}{$\langle\cos\varphi\rangle= 0.256$}.
\textcolor{black}{Using the renormalization $\langle\cos\varphi\rangle$ as a free parameter in the fitting is a  simple, phenomenological way to emphasize that $\langle\cos\varphi\rangle\neq 0$, which is unexpected for small Josephson junctions in series with a large resistance $(R\gg R_Q)$ when considering the DC behaviour. Regardless of the value used,} these \textcolor{black}{oscillations} indicate that the photonic heat exchange \textcolor{black}{from the drain to the source, i.e., current noise over a bandwidth $0-k_{\rm B}T/h\sim 4$ GHz, is mainly transmitted through the Josephson inductor channel, which acts as a low-pass filter}. At $\Phi = \Phi_0/2$ (dashed purple and dark red lines), the power is reduced as expected, in fair agreement with the data. In this regime, the Josephson critical current is vanishingly small (making the inductance essentially infinite at the relevant frequencies); consequently, the power transmitted from the drain to the source takes place mainly through the junction capacitance $C_\mr{J}$, which acts as a high-pass filter in the transmission and thus only enables a small fraction of the thermal fluctuations to be transmitted as current in the circuit. \textcolor{black}{The calculation was performed using an asymmetry parameter $d= 0.15$ (which does not coincide with that obtained for replica I, but matches much better the closed SQUID data in heat transport) for sample I  and is coincidentally the same for sample II}. Furthermore, the heat current calculated within the \textit{P(E)} theory at $\Phi = 0$ shows a decrease of $\dot{Q}_\nu$ to the level of the power obtained in the linear model at $\Phi = \Phi_0/2$ (see Supplementary Figs. \textcolor{black}{S5b and S5c}).

Figure \ref{Conductance}d and Fig. \ref{Conductance}e show the heat current modulations measured at given phonon temperatures $T_\mr{0}$ for samples I and II, respectively. Clear oscillations with period $\Phi_0$ are observed. This result unequivocally demonstrates that the inductive response of the junction persists in the presence of strong environmental back-action, in contrast to what is observed for charge transport measurements in the Replica. The data is again compared to the theoretical models proposed. 
\textcolor{black}{On the one hand, for the two examples presented, the heat current modulations are qualitatively captured with the linear model if we use a value of \textcolor{black}{$\langle\cos\varphi\rangle= 0.69$ for sample I and $\langle\cos\varphi\rangle= 0.256$ for the modulations at $T_\mr0 = 150$ mK, and $\langle\cos\varphi\rangle= 0.92$ at $T_\mr0 = 180$ mK in sample II}, \textcolor{black}{and keeping the asymmetry parameter same as before}.} 
On the other hand, the oscillation amplitude predicted by \textit{P(E)} theory (for which $\langle\cos\varphi\rangle\approx0$) is much smaller than the amplitudes observed (see Supplementary \textcolor{black}{Figs. S5d and S5e}).

Let us now focus on the discrepancy between charge and heat transport measurements. One obvious difference resides in the relevant frequency range: at zero frequency, $P(E)$ theory of Coulomb blockade describes incoherent Cooper pair tunneling through the junction, and the transition to an insulating state predicted by Schmid and Bulgadaev is observed as $R$ becomes greater than $R_\mr{Q}$. On the other hand, heat transport deals with non-zero frequency current fluctuations flowing through the junction at zero net voltage bias. This was considered previously \cite{Saira2016} through an extension of the static version of $P(E)$ theory to finite frequency transport. The derivation relies on the hypothesis of very weak Josephson coupling, \textcolor{black}{$2E_\mr{J} < k_\mr{B}T_\mr{S,D}$. This condition is not well satisfied for both samples. However, this point can hardly justify our contradicting observations: indeed, the discrepancy is the strongest for sample II (as highlighted by the sharp Coulomb gap observed in DC charge transport, see Fig. \ref{Replica}c), where we are closer to this limit}. An alternative, motivated by the microscopic description of the Josephson junction \textcolor{black}{in an arbitrary electromagnetic environment} \cite{joyez2013self}, is the existence of an inductive-like shunt in the junctions' environment, fundamentally due to the BCS gap, which protects the ground state of the junction from strong phase diffusion. The presence of such a shunt would translate as a finite supercurrent peak, which indeed was reported recently \cite{grimm2019bright}. Its absence in our charge measurement (and previous ones \cite{kuzmin1991coulomb,penttila1999superconductor}) contradicts this interpretation.

\textcolor{black}{Note that recent high-frequency measurements of a small Josephson junction in an engineered high impedance environment have revealed the inelastic nature of the scattering process of a photon off the junction \cite{kuzmin2021inelastic}. The conceptual similarity between the setup considered there and ours suggests that the description of the junction as a renormalized inductor \cite{joyez2013self}, while useful for a basic understanding, is too simplistic because the nonlinearities of the junctions are present due to strong phase fluctuations. Nevertheless, our bolometric technique collects photons at energies over a bandwidth $\sim k_\mr{B}T_\mr{0}/\hbar$ and, therefore, would not distinguish between several down-converted photons out of an inelastic process or a single elastically scattered photon.}

In summary, we have experimentally demonstrated through heat transport measurements that a Josephson junction acts as an inductor even in the presence of a highly resistive environment. Though the interpretation of the dissipative transition can be debated \cite{murani2020absence, Masuki2022}, the discrepancy between the heat transport measurements and the control charge transport measurements by us here and in previous works \cite{kuzmin1991coulomb, penttila1999superconductor, Herrero2002}, cannot be accounted for by the existing theory and calls for further developments, both experimental and theoretical. Our findings are important not only from the fundamental physics point of view but also for future applications such as microbolometers or heat sink designs in quantum circuits. On a practical side, we note that any design aiming at increasing resistances for improved, quantum-limited tunable remote electronic cooling \cite{MatthiasMeschkeandPekolaJukka2006,Timofeev2009} is much less sensitive to back-action effects than initially anticipated. 

\section{Methods}
\subsection*{Device fabrication and measurement}

The devices were fabricated on 4-inch silicon substrates covered by $300$ nm of Si/SiO$_2$ in an electron beam lithography (EBL, Vistec EBPG500 + operating at $100$ kV) using a Ge-based hard mask process and the conventional shadow evaporation technique \cite{Dolan1977}. The silicon wafer was coated with 400 nm layers of poly(methylmethacrylate-methacrylate acid) P(MMA-MAA) resist spun for 1 min at 5500 rpm and baked at 180 $^\circ$C for 20 min, twice. Then, on top of it 22 nm Ge layer was deposited in an electron-beam evaporator, and right after, approximately 50 nm thick of PMMA was coated with spun at 2500 rpm for 1 min and baked at 160 $^\circ$C for 1 min. The devices were patterned on the PMMA layer by using electron beam lithography, and afterward, it was developed using a mixture solution with a concentration of 1:3 of methyl-isobutyl-ketone+isopropanol (MIBK). This pattern is transferred to the Ge mask using reactive ion etching (RIE) with tetrafluoromethane CF$_\mr{4}$ plasma. The undercut in the MMA resist was created  by oxygen plasma in the same RIE chamber. The metallic parts were made in three evaporation steps: first, a 20 nm layer of Al is evaporated at an evaporation angle of $-22^\circ$. Then, static oxidation \textit{in-situ} with pressure around 3 mbar for 3 minutes is made. This step defines the superconducting finger used as a thermometer,  heater, and branch of the SQUID. In the second step, a 20 nm layer of Al is evaporated at an angle of $-7^\circ$, forming the SQUID and the clean superconducting contact. Finally, a 14 nm layer of Cr is evaporated at an angle of $24^\circ$ comprising the thermal bath. The nominal loop area of the SQUID for the two samples was 25 $\mu$m$^\mr{2}$. The main difference between them lies in the overlap area of the Josephson junction (JJ), which for Sample I is nominally 130$\times$140 nm$^\mr{2}$ and for Sample II is 85$\times$85 nm$^\mr{2}$. The resist was lifted-off in acetone at $52^\circ$C. Then, the sample is attached to a sample carrier to be electrically connected to it by Al wire bonds for being measured. 
The bonded sample is placed on a stage with a double brass enclosure that acts as a radiation shield. It is connected to the mixing chamber of a custom-made plastic dilution refrigerator with a base temperature of approximately 40 mK.  DC signals were applied through cryogenic signal lines filtered with lossy coaxial cables with 0-10 kHz bandwidth connected to the bonded sample through a room-temperature breakout box. 
In order to sweep the SQUID Josephson energy, a perpendicular magnetic field is supplied by applying  DC current to an external superconducting magnet inserted around the vacuum can. All the input signals were applied and read out using programmable sources and multimeters. Amplifying current and voltage output signal was accomplished using a room temperature low noise current amplifier Femto \text{DDPCA-300} and voltage amplifier Femto \text{DLVPA-100-F-D}, respectively. The cryostat temperature is controlled by applying a voltage across the heater resistance attached to the mixing chamber. The calibration of thermometers was done by monitoring the voltage drop across the SINIS configuration (current biased $I_\mr{th} = 15$ pA) at zero heating bias voltage while varying the cryostat temperature up to 500 mK \cite{Giazotto2006}.

\subsection*{Photon transmission coefficient}

As mentioned in the main text, the transmission probability of the thermal radiation from the source and drain $\tau(\omega, \Phi)$ used is Eq. (\ref{Landauer}) has been calculated within the two models. In the linear model approximation, $\tau(\omega,\Phi)$ can be written as

\begin{equation}
    \tau(\omega,\Phi)= \frac{4R_\mr{S}R_\mr{D}}{|Z_\mr{T}(\omega,\Phi)|^2},
\end{equation}
\noindent with 
\begin{equation}
 Z_\mr{T}(\omega,\Phi)= R_\mr{S}+R_\mr{D} + \frac{1}{-i\omega C_\mr{J} + \frac{2\pi I_\mr{c}}{-i\omega\Phi_0} \langle\cos{\varphi}\rangle}.
\end{equation}

In the charge dominated regime ($E_\mr{C}\gg k_\mr{B} T_\mr{S,D}\gg 2E_\mr{J}$) and taking into account the effect of the environment resistors through \textit{P(E)} function, the transmission probability $\tau(\omega,\Phi)$ for the system studied reads \cite{thomas2019photonic},

\begin{equation} \label{Transmision}
\begin{split}
\tau(\omega,\Phi) &= \frac{4R_\mr{S}R_\mr{D}}{\Bigg|R_\mr{S} + R_\mr{D} + \frac{1}{-i\omega C_\mr{J} + \tilde{Z}^{-1}(\omega,\Phi)}\Bigg|^\mr{2}}  \\
&+\frac{\pi^2 I_\mr{C}^\mr{2}}{2e^2}[P_\mr{S}(\omega)-P_\mr{S}(-\omega)][P_\mr{D}(\omega)-P_\mr{D}(-\omega)], 
\end{split}
\end{equation} 

\noindent where $\tilde{Z}(\omega,\Phi)$ is the effective frequency-dependent impedance and, the functions $P_\mr{S}(\omega)$ and $P_\mr{D}(\omega)$ represent the probability of photon absorption in the source and the drain resistors, respectively. These functions are defined as

\begin{equation}
P_\mr{l} (\omega) = \int\frac{dt}{2\pi}e^{i\omega t}e^{-J_\mr{l}(t)} ,
\end{equation}

\noindent here l = S, D and  $J_\mr{l}$ is the phase-phase correlation function given by \cite{ingold1992charge}:

\begin{equation}
J_\mr{l}(\omega) = \frac{4e^2R_\mr{l}}{\pi\hbar}\int_{0}^{\infty}d\omega \frac{\coth \frac{\hbar\omega}{2k_\mr{B}T_\mr{l}}(1-\cos\omega t) + i\sin \omega t}{\omega(1 + \omega^\mr{2}(R_\mr{S}+R_\mr{D})^\mr{2}C_\mr{J}^\mr{2})}.
\end{equation}

The effective impedance $\tilde{Z}(\omega,\Phi)$ is defined as 

\begin{equation}
\begin{split}
\frac{1}{\tilde{Z}(\omega,\Phi)} = \frac{\pi I_\mr{c}^\mr{2}}{2\hbar\omega}\big[ P(\omega) - P(-\omega) - \\ 
i(P(\omega) + P(-\omega) -2P(0))\tan \frac{\pi(R_\mr{S}+ R_\mr{D})}{R_\mr{Q}}  \big],
\end{split}
\end{equation}

\noindent where $P(\omega)$ is the \textit{P}-function of the effective environment defined by the convolution of the \textit{P(E)}-function of the two resistors,

\begin{equation}
P(\omega) = \int d\omega' P_\mr{S}(\omega - \omega') P_\mr{D}(\omega').
\end{equation} 

\section{Data availability}

The findings of this study can be supported with data that is accessible upon a reasonable request from the corresponding author.
\section{References}

\begin{thebibliography}{99}
\expandafter\ifx\csname url\endcsname\relax
  \def\url#1{\texttt{#1}}\fi
\expandafter\ifx\csname urlprefix\endcsname\relax\def\urlprefix{URL }\fi
\providecommand{\bibinfo}[2]{#2}
\providecommand{\eprint}[2][]{\url{#2}}

\bibitem{nyquist1928thermal}
\bibinfo{author}{Nyquist, H}.
\newblock \bibinfo{title}{Thermal agitation of electric charge in conductors}.
\newblock \emph{\bibinfo{journal}{Phys. Rev.}}
  \textbf{\bibinfo{volume}{32}}, \bibinfo{pages}{110}
  (\bibinfo{year}{1928}).

\bibitem{johnson1928thermal}
\bibinfo{author}{Johnson, J.~B}.
\newblock \bibinfo{title}{Thermal agitation of electricity in conductors}.
\newblock \emph{\bibinfo{journal}{Phys. Rev.}}
  \textbf{\bibinfo{volume}{32}}, \bibinfo{pages}{97}
  (\bibinfo{year}{1928}).

\bibitem{Schmidt2004}
\bibinfo{author}{Schmidt, D.~R.,  Schoelkopf, R.~J. \& Cleland, A.~N.} 
\newblock \bibinfo{title}{Photon-mediated thermal relaxation of electrons in nanostructures}.
\newblock \emph{\bibinfo{journal}{Phys. Rev. Lett.}}
  \textbf{\bibinfo{volume}{93}}, \bibinfo{pages}{045901}
  (\bibinfo{year}{2004}).

\bibitem{MatthiasMeschkeandPekolaJukka2006}
\bibinfo{author}{Meschke, M., Guichard, W. \& Pekola, J. P}.
\newblock \bibinfo{title}{Single-mode heat conduction by photons}.
\newblock \emph{\bibinfo{journal}{Nature.}}
  \textbf{\bibinfo{volume}{444}}, \bibinfo{pages}{187--190}
  (\bibinfo{year}{2006}).

\bibitem{partanen2018flux}
\bibinfo{author}{Partanen, M.} \emph{et~al.}
\newblock \bibinfo{title}{Flux-tunable heat sink for quantum electric circuits}.
\newblock \emph{\bibinfo{journal}{Sci. Rep.}}
\textbf{\bibinfo{volume}{8}}, \bibinfo{pages}{1--9}
(\bibinfo{year}{2018}).  

\bibitem{Ronzani2018}
\bibinfo{author}{Ronzani, A.} \emph{et~al.}
\newblock \bibinfo{title}{Tunable photonic heat transport in a quantum heat valve}.
\newblock \emph{\bibinfo{journal}{Nat. Phys.}}
\textbf{\bibinfo{volume}{14}}, \bibinfo{pages}{991--995}
(\bibinfo{year}{2018}).  

\bibitem{maillet2020electric}
\bibinfo{author}{Maillet, O.} \emph{et~al.}
\newblock \bibinfo{title}{Electric field control of radiative heat transfer in a superconducting circuit}.
\newblock \emph{\bibinfo{journal}{Nat. Commun.}}
\textbf{\bibinfo{volume}{11}}, \bibinfo{pages}{1--6}
(\bibinfo{year}{2020}).  

\bibitem{Ojanen2008}
\bibinfo{author}{Ojanen, T. \& Jauho, A.-P}.
\newblock \bibinfo{title}{Mesoscopic photon heat transistor}.
\newblock \emph{\bibinfo{journal}{Phys. Rev. Lett.}}
\textbf{\bibinfo{volume}{100}}, \bibinfo{pages}{155902}
(\bibinfo{year}{2008}).  

\bibitem{Pascal2011}
\bibinfo{author}{Pascal, L.~M.~A, Courtois, H. \& Hekking, F.~W.~J}.
\newblock \bibinfo{title}{Circuit approach to photonic heat transport}.
\newblock \emph{\bibinfo{journal}{Phys. Rev. B.}}
\textbf{\bibinfo{volume}{83}}, \bibinfo{pages}{125113}
(\bibinfo{year}{2011}).  

\bibitem{kuzmin2019quantum}
\bibinfo{author}{Kuzmin, R.} \emph{et~al.}
\newblock \bibinfo{title}{Quantum electrodynamics of a superconductor-insulator phase transition}.
\newblock \emph{\bibinfo{journal}{Nat. Phys.}}
\textbf{\bibinfo{volume}{15}}, \bibinfo{pages}{930--934}
(\bibinfo{year}{2019}).  

\bibitem{thomas2019photonic}
\bibinfo{author}{Thomas, G., Pekola, J.~P. \& Golubev, D.~S}.
\newblock \bibinfo{title}{Photonic heat transport across a {J}osephson junction}.
\newblock \emph{\bibinfo{journal}{Phys. Rev. B.}}
\textbf{\bibinfo{volume}{100}}, \bibinfo{pages}{094508}
(\bibinfo{year}{2019}).  

\bibitem{leger2019observation}
\bibinfo{author}{L{\'e}ger, S.} \emph{et~al.}
\newblock \bibinfo{title}{Observation of quantum many-body effects due to zero point fluctuations in superconducting circuits}.
\newblock \emph{\bibinfo{journal}{Nat. Commun.}}
\textbf{\bibinfo{volume}{10}}, \bibinfo{pages}{1--8}
(\bibinfo{year}{2019}).  

\bibitem{ingold1992charge}
\bibinfo{author}{Ingold, G.-L. \& Nazarov, Y.~V}.
\newblock \bibinfo{title}{Charge tunneling rates in ultrasmall junctions}.
\newblock (\bibinfo{publisherl}{Springer})
\bibinfo{pages}{21--107}
(\bibinfo{year}{1992}).  

\bibitem{kuzmin1991coulomb}
\bibinfo{author}{Kuzmin, L.~S} \emph{et~al.}
\newblock \bibinfo{title}{Coulomb blockade and incoherent tunneling of {C}ooper pairs in ultrasmall junctions affected by strong quantum fluctuations}.
\newblock \emph{\bibinfo{journal}{Phys. Rev. Lett.}}
\textbf{\bibinfo{volume}{67}}, \bibinfo{pages}{1161}
(\bibinfo{year}{1991}).  

\bibitem{averin1990incoherent}
\bibinfo{author}{Averin, D.~V., Nazarov, Y.~V. \& Odintsov, A.~A}.
\newblock \bibinfo{title}{Incoherent tunneling of the {C}ooper pairs and magnetic flux quanta in ultrasmall {J}osephson junctions}.
\newblock \emph{\bibinfo{journal}{Phys. B Condens. Matter.}}
\textbf{\bibinfo{volume}{165}}, \bibinfo{pages}{945--946}
(\bibinfo{year}{1990}).  

\bibitem{yagi1997phase}
\bibinfo{author}{Yagi, R., Kobayashi, S., and Ootuka, Y.}.
\newblock \bibinfo{title}{Phase diagram for superconductor-insulator transition in single small {J}osephson junctions with shunt resistor}.
\newblock \emph{\bibinfo{journal}{JPSJ}}
\textbf{\bibinfo{volume}{66}}, \bibinfo{pages}{3722--3724}
(\bibinfo{year}{1997}).  

\bibitem{penttila2001experiments}
\bibinfo{author}{Penttil{\"a}, J } \emph{et~al.}
\newblock \bibinfo{title}{Experiments on dissipative dynamics of single {J}osephson junctions}.
\newblock \emph{\bibinfo{journal}{J. Low Temp. Phys.}}
\textbf{\bibinfo{volume}{125}}, \bibinfo{pages}{89--114}
(\bibinfo{year}{2001}).  

\bibitem{corlevi2006phase}
\bibinfo{author}{Corlevi, S.} \emph{et~al.}
\newblock \bibinfo{title}{Phase-charge duality of a {J}osephson junction in a fluctuating electromagnetic environment}.
\newblock \emph{\bibinfo{journal}{Phys. Rev. Lett.}}
\textbf{\bibinfo{volume}{97}}, \bibinfo{pages}{096802}
(\bibinfo{year}{2006}).  

\bibitem{grimm2019bright}
\bibinfo{author}{Grimm, A.} \emph{et~al.}
\newblock \bibinfo{title}{Bright on-demand source of antibunched microwave photons based on inelastic {C}ooper pair tunneling}.
\newblock \emph{\bibinfo{journal}{Phys. Rev. X.}}
\textbf{\bibinfo{volume}{9}}, \bibinfo{pages}{021016}
(\bibinfo{year}{2019}).  

\bibitem{zhang2022suppressing}
\bibinfo{author}{Zhang, S.} \emph{et~al.}
\newblock \bibinfo{title}{Suppressing {A}ndreev bound state zero bias peaks using a strongly dissipative lead}.
\newblock \emph{\bibinfo{journal}{Phys. Rev. Lett.}}
\textbf{\bibinfo{volume}{128}}, \bibinfo{pages}{076803}
(\bibinfo{year}{2022}).  

\bibitem{schmid1983diffusion}
\bibinfo{author}{Schmid, A.}
\newblock \bibinfo{title}{Diffusion and localization in a dissipative quantum system}.
\newblock \emph{\bibinfo{journal}{Phys. Rev. Lett.}}
\textbf{\bibinfo{volume}{51}}, \bibinfo{pages}{1506}
(\bibinfo{year}{1983}).  


\bibitem{bulgadaev1984phase}
\bibinfo{author}{Bulgadaev, S.~A.}
\newblock \bibinfo{title}{Phase diagram of a dissipative quantum system}.
\newblock \emph{\bibinfo{journal}{JETP Lett.}}
\textbf{\bibinfo{volume}{39}}, \bibinfo{pages}{264--267}
(\bibinfo{year}{1984}).  


\bibitem{penttila1999superconductor}
\bibinfo{author}{Penttil{\"a}, J.~S.} \emph{et~al.}
\newblock \bibinfo{title}{“{S}uperconductor-Insulator transition” in a single {J}osephson junction}.
\newblock \emph{\bibinfo{journal}{Phys. Rev. Lett.}}
\textbf{\bibinfo{volume}{82}}, \bibinfo{pages}{1004}
(\bibinfo{year}{1999}).  


\bibitem{murani2020absence}
\bibinfo{author}{Murani, A.} \emph{et~al.}
\newblock \bibinfo{title}{Absence of a dissipative quantum phase transition in {J}osephson junctions}.
\newblock \emph{\bibinfo{journal}{Phys. Rev. X.}}
\textbf{\bibinfo{volume}{10}}, \bibinfo{pages}{021003}
(\bibinfo{year}{2020}).  

\bibitem{Perti2020}
\bibinfo{author}{Hakonen, P.~J. \& Sonin, E.~B.}
\newblock \bibinfo{title}{Comment on ``Absence of a Dissipative Quantum Phase Transition in {J}osephson Junctions''}.
\newblock \emph{\bibinfo{journal}{Phys. Rev. X.}}
\textbf{\bibinfo{volume}{11}}, \bibinfo{pages}{018001}
(\bibinfo{year}{2020}).  


\bibitem{JoyezReply}
\bibinfo{author}{Murani, A.} \emph{et~al.}
\newblock \bibinfo{title}{Reply to ``Comment on `Absence of a Dissipative Quantum Phase Transition in {J}osephson Junctions' ''}.
\newblock \emph{\bibinfo{journal}{Phys. Rev. X.}}
\textbf{\bibinfo{volume}{11}}, \bibinfo{pages}{018002}
(\bibinfo{year}{2021}).  

\bibitem{Masuki2022}
\bibinfo{author}{Masuki, K.} \emph{et~al.}
\newblock \bibinfo{title}{Absence versus Presence of Dissipative Quantum Phase Transition in {J}osephson Junctions}.
\newblock \emph{\bibinfo{journal}{Phys. Rev. Lett.}}
\textbf{\bibinfo{volume}{129}}, \bibinfo{pages}{087001}
(\bibinfo{year}{2022}).  


\bibitem{Sepulcre2022}
\bibinfo{author}{Sépulcre, T., Florens, S. \& Snyman, I.}
\newblock \bibinfo{title}{Comment on ``Absence versus Presence of Dissipative Quantum Phase Transition in {J}osephson Junctions''}.
\newblock \bibinfo{journal}{Preprint at 
https://doi.org/10.48550/arXiv.2210.00742 (2022)}.


\bibitem{MasukiReply2022}
\bibinfo{author}{Masuki, K.} \emph{et~al.}
\newblock \bibinfo{title}{Reply to  'Comment on ``Absence versus Presence of dissipative quantum phase Transition in {J}osephson Junctions''}.
\newblock \bibinfo{journal}{Preprint at 
https://doi.org/10.48550/arXiv.2210.10361
 (2022)}.


\bibitem{kuzmin2023observation}
\bibinfo{author}{Kuzmin, R.}\emph{et~al.}
\newblock \bibinfo{title}{Observation of the {S}chmid-{B}ulgadaev dissipative quantum phase transition}.
\newblock \bibinfo{journal}{Preprint at 
https://doi.org/10.48550/arXiv.2304.05806 (2023)}.



\bibitem{giazotto2012josephson}
\bibinfo{author}{Giazotto, F. \& Mart{\'\i}nez-P{\'e}rez, M.~J.}
\newblock \bibinfo{title}{The {J}osephson heat interferometer}.
\newblock \emph{\bibinfo{journal}{Nature.}}
\textbf{\bibinfo{volume}{492}}, \bibinfo{pages}{401--405}
(\bibinfo{year}{2012}).  


\bibitem{Sivre2018}
\bibinfo{author}{Sivre, E.} \emph{et~al.}
\newblock \bibinfo{title}{Heat {C}oulomb blockade of one ballistic channel}.
\newblock \emph{\bibinfo{journal}{Nat. Phys.}}
\textbf{\bibinfo{volume}{14}}, \bibinfo{pages}{145--148}
(\bibinfo{year}{2018}).  

\bibitem{Timofeev2009}
\bibinfo{author}{Timofeev, A.} \emph{et~al.}
\newblock \bibinfo{title}{Electronic refrigeration at the quantum limit}.
\newblock \emph{\bibinfo{journal}{Phys. Rev. Lett.}}
\textbf{\bibinfo{volume}{102}}, \bibinfo{pages}{200801}
(\bibinfo{year}{2009}).  

\bibitem{andreev1965thermal}
\bibinfo{author}{Andreev, A.~F.}
\newblock \bibinfo{title}{Thermal conductivity of the intermediate state of superconductors {II}}.
\newblock \emph{\bibinfo{journal}{Sov. Phys. JETP.}}
\textbf{\bibinfo{volume}{20}}, \bibinfo{pages}{1490}
(\bibinfo{year}{1965}).  

\bibitem{Giazotto2006}
\bibinfo{author}{Giazotto, F.} \emph{et~al.}
\newblock \bibinfo{title}{Opportunities for mesoscopic in thermometry and refrigeration: Physics and applications}.
\newblock \emph{\bibinfo{journal}{Rev. Mod. Phys.}}
\textbf{\bibinfo{volume}{78}}, \bibinfo{pages}{217}
(\bibinfo{year}{2006}).  

\bibitem{devoret1990effect}
\bibinfo{author}{Devoret, M.} \emph{et~al.}
\newblock \bibinfo{title}{Effect of the electromagnetic environment on the {C}oulomb blockade in ultrasmall tunnel junctions}.
\newblock \emph{\bibinfo{journal}{Phys. Rev. Lett.}}
\textbf{\bibinfo{volume}{64}}, \bibinfo{pages}{1824}
(\bibinfo{year}{1990}).  

\bibitem{lu2023phase}
\bibinfo{author}{Lu, Wen-Sen} \emph{et~al.}
\newblock \bibinfo{title}{Phase diffusion in low-$E_\mr{J}$ Josephson junctions at millikelvin temperatures}.
\newblock \emph{\bibinfo{journal}{Electronics}}
\textbf{\bibinfo{volume}{12}}, \bibinfo{pages}{416}
(\bibinfo{year}{2023}).  

\bibitem{pendry1983quantum}
\bibinfo{author}{Pendry, J.}
\newblock \bibinfo{title}{Quantum limits to the flow of information and entropy}.
\newblock \emph{\bibinfo{journal}{J. Phys. A Math: Mathematical and General.}}
\textbf{\bibinfo{volume}{16}}, \bibinfo{pages}{2161}
(\bibinfo{year}{1983}).  

\bibitem{joyez2013self}
\bibinfo{author}{Joyez, P.}
\newblock \bibinfo{title}{Self-consistent dynamics of a {J}osephson junction in the presence of an arbitrary environment}.
\newblock \emph{\bibinfo{journal}{Phys. Rev. Lett.}}
\textbf{\bibinfo{volume}{110}}, \bibinfo{pages}{217003}
(\bibinfo{year}{2013}).  

\bibitem{pechenezhskiy2020superconducting}
\bibinfo{author}{Pechenezhskiy, I., }\emph{et~al.}.
\newblock \bibinfo{title}{The superconducting quasicharge qubit}.
\newblock \emph{\bibinfo{journal}{Nature.}}
  \textbf{\bibinfo{volume}{585}}, \bibinfo{pages}{368--371}
  (\bibinfo{year}{2020}).


\bibitem{Saira2016}
\bibinfo{author}{Saira, O.~P.} \emph{et~al.}
\newblock \bibinfo{title}{Dispersive Thermometry with a {J}osephson Junction Coupled to a Resonator}.
\newblock \emph{\bibinfo{journal}{Phys. Rev. Applied.}}
\textbf{\bibinfo{volume}{6}}, \bibinfo{pages}{024005}
(\bibinfo{year}{2016}).  

\bibitem{kuzmin2021inelastic}
\bibinfo{author}{Kuzmin, R.} \emph{et~al.}
\newblock \bibinfo{title}{Inelastic Scattering of a Photon by a Quantum Phase Slip}.
\newblock \emph{\bibinfo{journal}{Phys. Rev. Lett.}}
\textbf{\bibinfo{volume}{126}}, \bibinfo{pages}{197701}
(\bibinfo{year}{2021}).  


\bibitem{Herrero2002}
\bibinfo{author}{Herrero, C.~P. \& Zaikin, A.~D.}
\newblock \bibinfo{title}{Superconductor-insulator quantum phase transition in a single {J}osephson junction}.
\newblock \emph{\bibinfo{journal}{Phys. Rev. B.}}
\textbf{\bibinfo{volume}{65}}, \bibinfo{pages}{104516}
(\bibinfo{year}{2002}).  


\bibitem{Dolan1977}
\bibinfo{author}{Dolan, G.~J.}
\newblock \bibinfo{title}{Offset masks for lift-off photoprocessing}.
\newblock \emph{\bibinfo{journal}{Appl. Phys. Lett.}}
\textbf{\bibinfo{volume}{31}}, \bibinfo{pages}{337--339}
(\bibinfo{year}{1977}).  




  

\end{thebibliography}

\section{Acknowledgements}
We thank P. Joyez, C. Altimiras, T. Yamamoto, J. Ankerhold, and J. Stockburger for illuminating discussions. We thank funding from Academy of Finland grant 336810, the Spanish State Research Agency through Grant RYC-2016-20778, and the EU for FET-Open contract AndQC. D.S. and J.P.P. acknowledge the support from the innovation program under the European Research Council (ERC) program (grant agreement 742559). This research was achieved using Otaniemi Research Infrastructure for Micro and Nanotechnologies (OtaNano).

\section{Author contributions}

The experiment was conceived by D.S., O.M., and J.P.P. and carried out by D.S. with contribution from O.M. and technical support by J.T.P. Sample fabrication was made by D.S. The theoretical model for heat transport based on the \textit{P(E)} theory was proposed by D.S.G, and D.S. performed the simulations. The data were analyzed, and the manuscript was written by D.S. with important contributions from all the authors. 

\section{Competing interests}

The authors declare no competing interests.

\end{document}


\title{Supplementary information: Bolometric detection of Josephson inductance in a highly resistive environment}
	\author{Diego Subero}\email{diego.suberorengel@aalto.fi}
	\affiliation{PICO Group, QTF Centre of Excellence, Department of Applied Physics, Aalto University School of Science, P.O. Box 13500, 0076 Aalto, Finland}
	\author{Olivier Maillet}
	\affiliation{PICO Group, QTF Centre of Excellence, Department of Applied Physics, Aalto University School of Science, P.O. Box 13500, 0076 Aalto, Finland}
	\affiliation{Université Paris-Saclay, CEA, CNRS, SPEC, 91191 Gif-sur-Yvette, France}
	
	\author{Dmitry S. Golubev}
	\author{George Thomas}
	\author{Joonas T. Peltonen}
	
	\affiliation{PICO Group, QTF Centre of Excellence, Department of Applied Physics, Aalto University School of Science, P.O. Box 13500, 0076 Aalto, Finland}
	\author{Bayan Karimi}
	\affiliation{PICO Group, QTF Centre of Excellence, Department of Applied Physics, Aalto University School of Science, P.O. Box 13500, 0076 Aalto, Finland}
	\affiliation{QTF Centre of Excellence, Department of Physics, Faculty of Science, University of Helsinki, 00014 Helsinki, Finland}
	\author{Marco Marín-Suárez}
	
	\affiliation{PICO Group, QTF Centre of Excellence, Department of Applied Physics, Aalto University School of Science, P.O. Box 13500, 0076 Aalto, Finland}
	
	\author{Alfredo Levy Yeyati}
	
	\author{Rafael Sánchez}
	
	\author{Sunghun Park}
	
	\affiliation{Departamento de Física Teórica de la Materia Condensada, Condensed Matter Physics Center (IFIMAC) and Instituto Nicolás Cabrera, Universidad Autonoma de Madrid, 28049 Madrid, Spain}

	\author{Jukka P. Pekola}
	\affiliation{PICO Group, QTF Centre of Excellence, Department of Applied Physics, Aalto University School of Science, P.O. Box 13500, 0076 Aalto, Finland}

	\maketitle
	\widetext
	
	\textcolor{black}{\section{S1. Electron-phonon coupling measurement}}
	
	\textcolor{black}{In order to measure the electron-phonon coupling constant of chromium, we utilized the hot electron effect under steady-state conditions. The studied system is depicted in Supplementary Fig. \ref{ep}.a. The Cr- film (red) has nominally the same dimension as those placed in the main and Replica samples and evaporated with the same target. The electron temperature in the film metal is elevated by applying Joule power $P= IV$ while simultaneously measuring its electron temperature, see Supplementary Fig. \ref{ep}.b. Assuming that the quasiparticle heat flow through the Al contact is negligible at temperatures below approximately 300 mK due to the good thermal insulation of superconducting Al, the dominant cooling mechanism for electrons in the heated film is electron-phonon scattering. Therefore, under steady-state conditions, we obtain}
	
	\begin{equation}\label{Power}
	IV = I^2 R_{\mr e}= \Sigma_n \Omega (T_\mr e^{n}- T_\mr 0^{n}),
	\end{equation}
	
	\noindent  \textcolor{black}{with $\Sigma_n$ the electron-phonon coupling, $\Omega$ the volume of the Cr-strip. Here, $\Sigma$ and $n$ are fitting parameters at each bath temperature, and their behaviors are shown in Supplementary Fig. \ref{ep}.c.  Additionally, we measure the resistance of the Cr-strip by fitting the IV curve, resulting to be $R_\mr e = 11$ k$\Omega$.} 
	
	\begin{figure*}[ht!]
		\centering
		\includegraphics[width=1\textwidth]{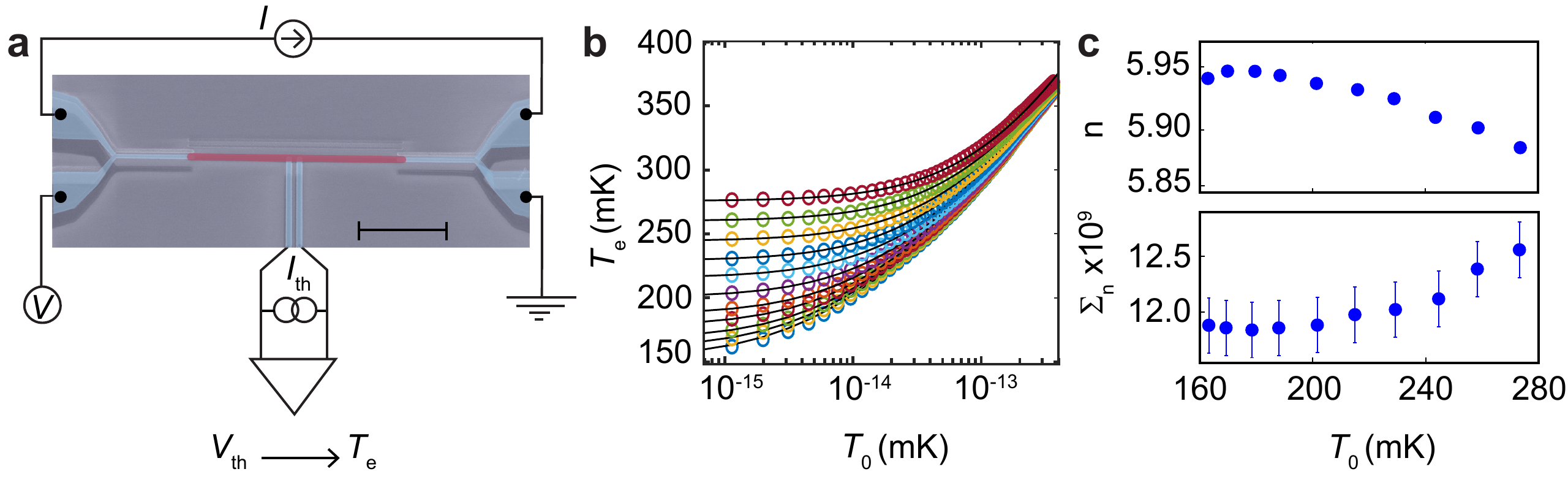}
		\caption{\textcolor{black}{\textbf{Experimental setup for measuring the electron-phonon coupling.} a.- Colored scanning electron micrograph (scale bar: 5 $\mu$m) of chromium (red) used to obtain the electron-phonon coupling, along with a schematic circuit diagram. Vertical aluminum leads are connected to the Cr-strip through an oxide tunnel barrier to monitoring its electronic temperature. b.- Electron temperature of Cr-strip at different $T_0$ as a function of the applied heating power, with black lines representing fitting results obtained by Supplementary Eq. (\ref{Power}). c.- Temperature dependence of the exponent $n$ and $\Sigma_n$ (in WK$^{-n}$m$^{-3}$) obtained by fitting measurements to Supplementary Eq. (\ref{Power})}.} \label{ep}
	\end{figure*}
	
	\textcolor{black}{\section{S2. Cleanliness of the contact between chromium and aluminum}}
	
	\textcolor{black}{An important assumption for our experiment's overall accuracy is negligible contact resistance between the Cr and Al films, reflecting efficient Andreev reflection as a way to prevent heat diffusion while allowing charge transport at the N/S interface. To confirm the cleanliness of the contact between Al and Cr, we measured various Cr-strips with different lengths but the same nominal transverse dimensions as those stated in Fig. 1a of the main text. The resistance scales linearly with the Cr strip length, as displayed in Supplementary Fig. \ref{scaling}, with an extrapolated residual resistance $\sim 376~\Omega$ that remains within the typical scatter encountered in the measurements. This value is much smaller than the resistance used in our experiment ($\sim 11~\mr k\Omega$), indicating a clean enough interface.} 
	
	\begin{figure*}[ht!]
		\centering
		\includegraphics[width=0.35\textwidth]{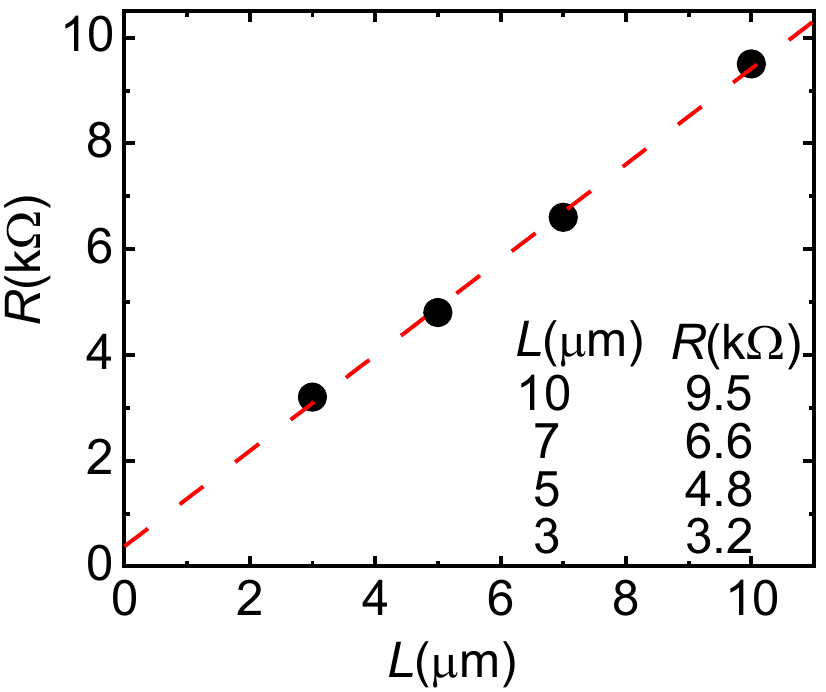}
		\caption{\textbf{Linear scaling of Cr resistance with the length.} Measured 4-wire resistances of Cr-strips with various lengths. The red line is a linear fit.} \label{scaling}
	\end{figure*}
	
	\section{S3. Current-voltage characteristic of the Replica sample}
	
	The quasiparticle tunnel resistance of the SQUID $R_\mr{J}$ and the superconducting gap were obtained from standard IV measurements as depicted in Supplementary Figs. \ref{IV}.a and \ref{IV}.b. The total resistance in series of the Replica $R_\mr{T}= R_\mr{S}+ R_\mr{D} + R_\mr{J}$ with $R_\mr S= R_\mr D= R_\mr e$,  was deduced from the slope of the IV-curve at high voltage bias and it is represented by the dashed black line in Supplementary Figs. \ref{IV}.a and \ref{IV}.b. These resistances were: 122 k$\Omega$ for Replica I and 280 k$\Omega$ for Replica II.
	Hence, the resistance $R_\mr{J}$ was determined by subtracting the effect of series resistance $R_\mr{S}$ and $R_\mr{D}$ from $R_\mr{T}$. Additionally, the superconducting gap was measured to be $\Delta\simeq 200\;\mu$eV. Thus, the Josephson energy is calculated using the Ambegaokar-Baratoff relation $E_\mr{J}= \Phi_0\Delta/4eR_\mr{J}$. The single charging energy of the junction $E_\mr{C}= e^2/2C_\mr{J}$ is extracted from the current-voltage curve at a low bias, as explained in the main text, from which we estimate the SQUID capacitance to be 1.5 fF for Replica I and 0.7 fF for Replica II.  
	
	\begin{figure*}[ht!]
		\centering
		\includegraphics[width=0.75\textwidth]{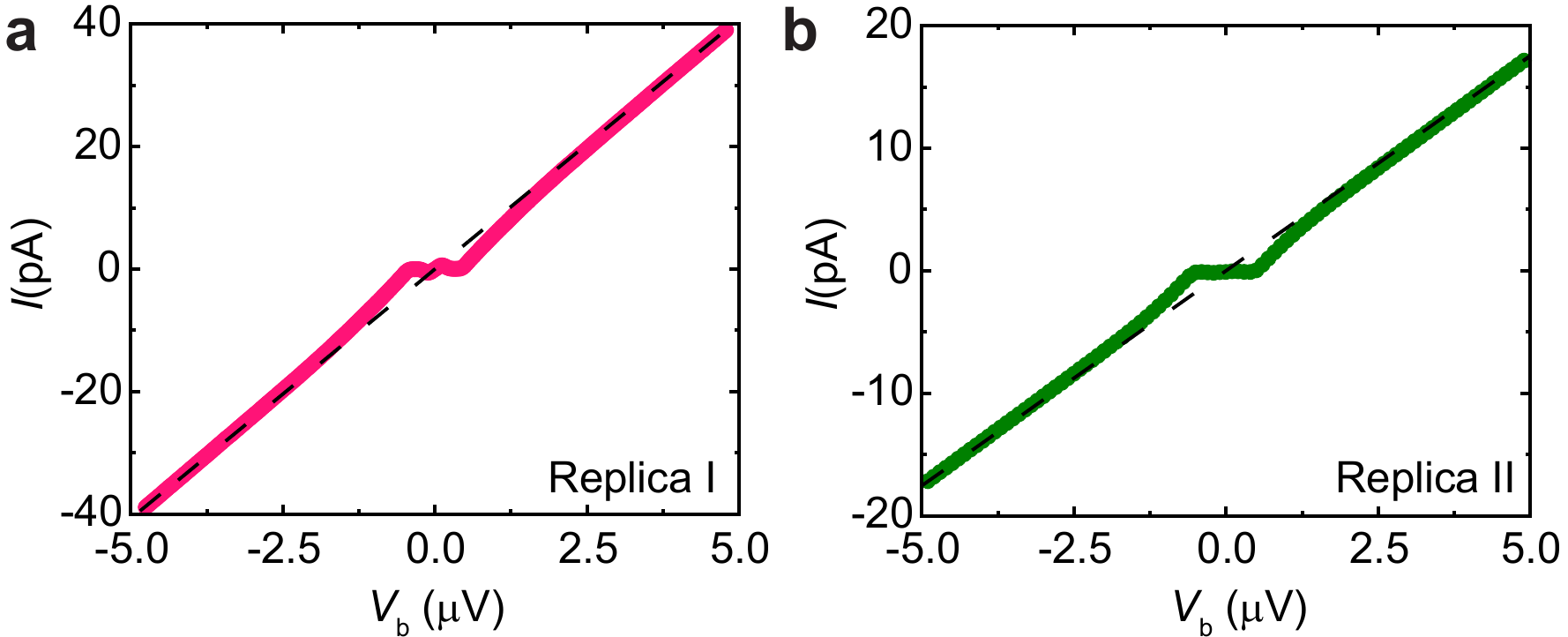}
		\caption{\textbf{Current-voltage measurements of the replica samples.} a, b.- Current-voltage characteristic at large voltage bias for the two Replica samples. The dashed black line is the linear fit.} \label{IV}
	\end{figure*}

	\section{S4. Josephson current of a small junction embedded in an electromagnetic environment}
	
	\begin{figure*}[ht!]
		\centering
		\includegraphics[width=1\textwidth]{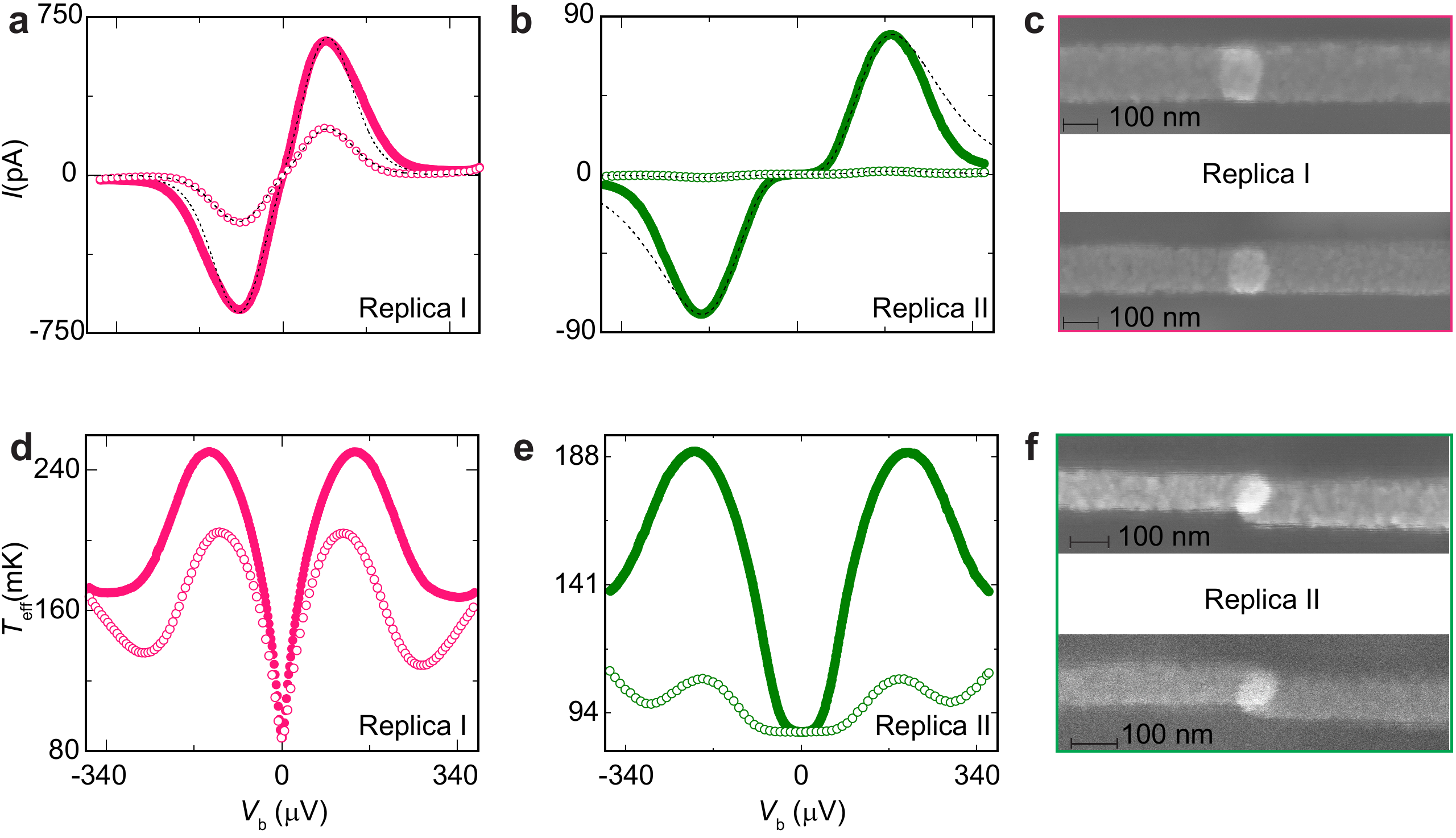}
		\caption{\textcolor{black}{\textbf{Enlargement of the IV curves shown in Supplementary Fig. \ref{IV} in the low voltage bias regime along with SEM images of the Josephson junction}. a, b.- Same data as shown in Fig. 2b and 2c in the main text. The dashed line displays the theoretical results obtained by using Supplementary Eq. (\ref{current}).
				d, e.-Bias dependence of the temperature of the resistors for the two Replicas in the low voltage bias regime at two different magnetic fluxes $\Phi = 0$ (solid circles) and $\Phi = \Phi_0/2$ (open circles), obtained from Supplementary Eq. (\ref{Temperature}). c, f.- SEM images of the two Josephson junctions of the SQUID of the Replicas.}} \label{heating}
	\end{figure*}
	
	\textcolor{black}{It is well established that the current-voltage characteristic of a Josephson junction with small critical current embedded in an electromagnetic environment at low voltage bias (shown in Supplementary Fig. \ref{heating}.a and  \ref{heating}.b ), is written as \cite{ingold1992charge, Saira2016} }

	\begin{equation}\label{current}
	I = \frac{\pi e E_\mr{J}^2(\Phi)}{\hbar}[P(2eV) - P(-2eV)],
	\end{equation}
	
	\noindent where $P(E)$ is the probability function that describes the energy exchange in the inelastic Cooper pair tunneling with the environment. This probability density is given by
	
	\begin{equation}\label{Pj}
	P(E) = \frac{1}{2\pi\hbar}\int e^{iEt}\langle e^{i\hat{\varphi}(t)} e^{-i\hat{\varphi}(0)}\rangle =  \frac{1}{2\pi\hbar}\int dt e^{i\frac{Et}{\hbar}-J(t)} ,
	\end{equation}
	
	\noindent where
	
	\begin{equation}\label{phase}
	J(t) = \frac{4e^2}{\pi\hbar}\int_{0}^{\infty} d\omega \operatorname{Re} [Z_\mr{T}(\omega)] \Big[\coth\frac{\hbar\omega}{2k_\mr{B}T_{\rm {eff}}(V_\mr{b})}\frac{1-\cos\omega t}{\omega} + i\frac{\sin\omega t}{\omega}\Big],
	\end{equation}
	
	\noindent is the phase-phase correlation function. Here, $Z_T(\omega)$ and $T_{\mr {eff}}(V)$ are the impedance seen by the junction and the bias voltage-dependent temperature of the resistors. The latter is modeled in the usual way,
	
	\begin{equation}\label{Temperature}
	T_{\mr{eff}}(V) = \Big[T_\mr{0}^{5.93} + \frac{IV_\mr{b}}{2\Sigma \Omega}\Big]^{1/5.93},
	\end{equation}
	
	\noindent and its behavior is shown in Supplementary Figs. \ref{heating}.d and \ref{heating}.e. Here, $T_0$ is the phonon temperature, $\Sigma= 12\times10^9$ WK$^{-5.93}$m$^{-3}$ is the electron-phonon constant of the normal metal, and $\Omega= 1.4\times10^{-20}$ m$^3$ is the volume of each resistor. Factor 2 in Supplementary Eq. (\ref{Temperature}) accounts for the two resistors surrounding the SQUID. The impedance $Z_\mr{T}(\omega)$ is derived from the resistively and capacitively shunted junction (RCSJ) model,
	
	\begin{equation}\label{impedanceT}
	Z_T(\omega) = \frac{1}{-i\omega C_\mr{J} + (R_\mr{S} + R_\mr{D})^{-1}} \ ,\; \operatorname{Re}[Z_T(\omega)] = \frac{R_\mr{S} + R_\mr{D}}{1 + \omega^2(R_\mr{S} + R_\mr{D})^2C_\mr{J}^{2}}\ .
	\end{equation}

	Note that the asymmetry of the critical current of the SQUID in  Supplementary Eq. (\ref{current}) is taken into account through the parameter \textit{d} as \cite{Ronzani2018}
	
	\begin{equation}
	E_\mr{J} (\Phi)=E_J(0) |\cos(\pi\Phi/\Phi_0)|\sqrt{1 + d^2\tan^2(\pi\Phi/\Phi_0)}.
	\end{equation}
	
	\textcolor{black}{The SEM pictures displayed in Supplementary Fig. \ref{heating}.c and Fig. \ref{heating}.f are used to assess the asymmetry of the SQUID for the two samples that would ideally be symmetric. However, the high peak in the IV curve at $\Phi= \Phi_0/2$ for Replica I (Supplementary Fig. \ref{heating}.a) suggests us to use a significant asymmetry parameter in fitting, which cannot be justified just based on the SEM pictures.}

	\section{S5. Heat current through a SQUID connected in series with resistors based on the \textit{P(E)}-theory}
	
	\begin{figure*}[ht!]
		\centering
		\includegraphics[width=1\textwidth]{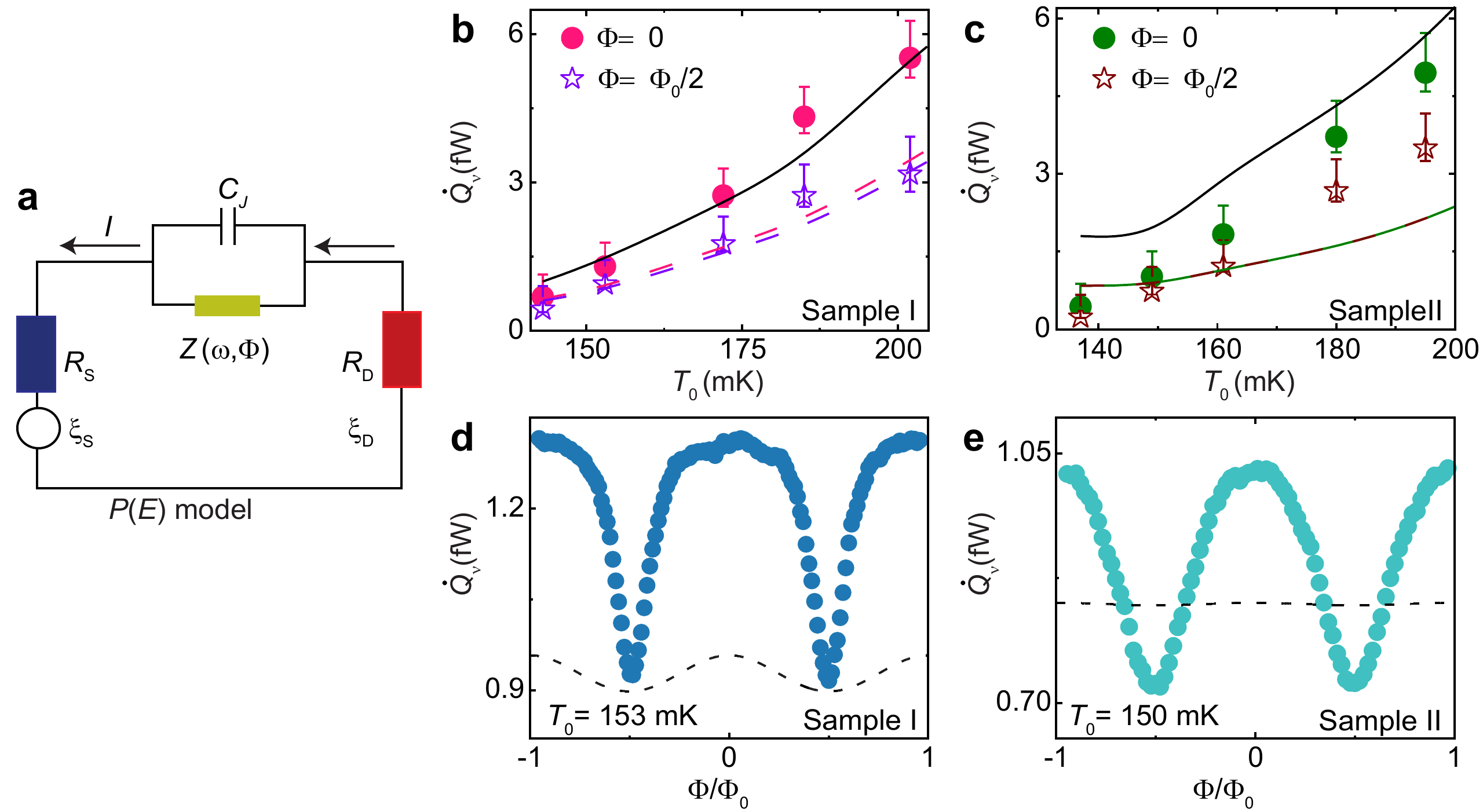}
		\caption{\textbf{$P(E)$-model for heat transport}. a.-Schematic electric circuit of the device with an effective impedance $\Tilde{Z}(\omega, \Phi)$ replacing the SQUID, in series with resistors. b, c, d, e- Same data as shown in Fig. 4 of the main text. The solid line represents the power transmitted through a single channel at the quantum limit $\dot{Q}_\mr{Q}$. The dashed lines are the theoretical results obtained from Supplementary Eq. (\ref{heat}) for the respective magnetic fluxes applied, with the circuit parameters obtained from fitting the IVC of the Replica samples in Fig. 2b and 2c, shown in the main text.} \label{Resistors}
	\end{figure*}
	
	Here, we consider the system shown in Supplementary Fig. \ref{Resistors}.a. The photonic heat flux between the two resistors can be expressed as (see Ref. \cite{thomas2019photonic})
	
	\begin{eqnarray}\label{heat}
	\dot{Q}_\mr{\nu} = \int_0^\infty \frac{d\omega}{2\pi}\,\hbar\omega\,\tau(\omega)
	\left[\frac{1}{e^{\hbar\omega/k_BT_\mr{D}}-1}-\frac{1}{e^{\hbar\omega/k_BT_\mr{S}}-1}\right]
	+  \frac{\pi\hbar I_C^2}{4e^2}\int_{-\infty}^{+\infty} d\omega\, \omega\, P_\mr{D}(\omega)P_\mr{S}(-\omega).
	\end{eqnarray}
	
	This equation is similar to the one deduced by Thomas. \textit{et.al.} \cite{thomas2019photonic} for a SQUID connected in parallel with resistors. However, one should re-define the photon transmission probability $\tau(\omega)$ and the functions $P_\mr{S}(\omega)$ and $P_\mr{D}(\omega)$ for the circuit shown in Supplementary Fig. \ref{Resistors}.a. Namely, the transmission probability is expressed as
	
	\begin{eqnarray}
	\tau(\omega) = \frac{4R_\mr{S}R_\mr{D}}{\left| R_\mr{S} + R_\mr{D} + \frac{1}{-i\omega C_\mr{J} + \frac{1}{\Tilde{Z}(\omega)}}\right|^2},
	\end{eqnarray}
	
	\noindent where the effective impedance of the junction $\Tilde{Z}(\omega)$ was already defined in the methods section of the main text (see  Eq. (8) in the main text). To find the expressions for the functions $P_\mr{j}(\omega)$ (j= S, D), we write down the classical equation of motion for the Josephson phase in the circuit of Supplementary Fig. \ref{Resistors}.a,
	
	\begin{eqnarray}
	C\frac{\hbar\ddot\varphi}{2e} + \frac{1}{R_\mr{S}+R_\mr{D}}\frac{\hbar\dot\varphi}{2e} + I_C\sin\varphi = \frac{R_\mr{S}\xi_\mr{S}+R_\mr{D}\xi_\mr{D}}{R_\mr{S}+R_\mr{D}}.
	\label{Eq1}
	\end{eqnarray}
	
	In order to obtain the equation for the phase, we introduce the new effective resistances ${\cal R}_\mr{S}$ and ${\cal R}_\mr{D}$ such that the Supplementary Eq. (\ref{Eq1}) reads,
	
	\begin{eqnarray}
	C_\mr{J}\frac{\hbar\ddot\varphi}{2e} + 
	\left(\frac{1}{{\cal R}_\mr{S}} + \frac{1}{{\cal R}_\mr{D}}\right)
	\frac{\hbar\dot\varphi}{2e} + I_\mr{C}(\Phi)\sin\varphi = \eta_\mr{S} + \eta_\mr{D}
	\label{Eq3}
	\end{eqnarray}
	
	\noindent where
	\begin{eqnarray}\label{rep}
	{\cal R}_\mr{j}= \frac{(R_\mr{S}+R_\mr{D})^2}{R_\mr{j}},\;\eta_\mr{j} = \frac{R_\mr{j}\xi_\mr{j}}{R_\mr{S}+R_\mr{D}},\;\;\; |\eta_\mr{j}|^2_\omega = \frac{R_\mr{j}}{(R_\mr{S}+R_\mr{D})^2}\omega\coth\frac{\hbar\omega}{2k_BT_\mr{j}}.
	\label{eta}
	\end{eqnarray}

	Thus, we find the functions $P_\mr{j}(\omega)$ by replacing the expression (\ref{rep}) in the Eqs. (\ref{Pj}, \ref{phase}) 
	
	\begin{eqnarray}
	P_\mr{j}(\omega) = \int\frac{dt}{2\pi}\, e^{i\omega t}\, e^{-J_\mr{j}(t)},
	\label{Pj1}
	\end{eqnarray}
	
	\noindent where
	\begin{eqnarray}
	J_j(t) &=& \frac{4e^2}{\pi\hbar}\int_0^\infty d\omega\, 
	\frac{\coth\frac{\hbar\omega}{2k_\mr{B}T_\mr{j}}(1-\cos\omega t) + i\sin\omega t}
	{{\cal R}_\mr{j}\left|-i\omega C_\mr{J} + \frac{1}{{\cal R}_\mr{S}} + \frac{1}{{\cal R}_\mr{D}}\right|^2}
	= \frac{4e^2}{\pi\hbar}\int_0^\infty d\omega\, 
	\frac{R_j\left[\coth\frac{\hbar\omega}{2k_\mr{B}T_\mr{j}}(1-\cos\omega t) + i\sin\omega t\right]}
	{(R_\mr{S}+R_\mr{D})^2\left|-i\omega C + \frac{R_\mr{S}}{(R_\mr{S}+R_\mr{D})^2} + \frac{R_\mr{D}}{(R_\mr{S}+R_\mr{D})^2}\right|^2}
	\nonumber\\
	&=& \frac{4e^2}{\pi\hbar}\int_0^\infty d\omega\, 
	\frac{R_\mr{j}\left[\coth\frac{\hbar\omega}{2k_\mr{B}T_\mr{j}}(1-\cos\omega t) + i\sin\omega t\right]}
	{(R_\mr{S}+R_\mr{D})^2\left|-i\omega C_\mr{J} + \frac{1}{R_\mr{S}+R_\mr{D}}\right|^2}
	= \frac{4e^2 R_\mr{j}}{\pi\hbar}\int_0^\infty d\omega\, 
	\frac{\coth\frac{\hbar\omega}{2k_\mr{B}T_\mr{j}}(1-\cos\omega t) + i\sin\omega t}{\left|1-i\omega (R_\mr{S}+R_\mr{D}) C_\mr{J} \right|^2}.
	\label{Jj1}
	\end{eqnarray}
	
	\section{S6. Impedance matching in the circuit}
	
	\textcolor{black}{To better understand the meaning of ``improved impedance matching" in our context, let us examine the net power flow between the two resistors described by Eq. (2) in the main text and repeated below,
		\begin{equation}
		\dot{Q}_\nu = \int_{0}^{\infty} \frac{d\omega}{2\pi}\hbar\omega\tau(\omega, \Phi)\left[ \frac{1}{e^{\frac{\hbar\omega}{k_BT_D}}-1}- \frac{1}{e^{\frac{\hbar\omega}{k_BT_S}}-1}\right]\ ,
		\end{equation} 
	}
	\noindent\textcolor{black}{where $\tau(\omega,\Phi)$ is the transmission coefficient for the power per unit frequency bandwidth from the drain resistor to the source resistor. This generic formula can be obtained with many approaches \cite{Schmidt2004, Ojanen2008, Pascal2011}. In a circuit description \cite{Schmidt2004}, which in our case is a simple lumped element one (see main text), $\tau(\omega,\Phi)= \frac{4R_SR_D}{\mid Z_{\rm T}(\omega, \Phi)\mid^2}$, where $Z_{\rm T}(\omega,\Phi)$ is the total impedance of the loop circuit. This can be expressed as,
		\begin{equation}
		Z_{\rm T}(\omega,\Phi) = R_{\rm S} + R_{\rm D} + Z_{\rm J}(\omega,\Phi) + Z_{\rm K}(\omega). \label{Z_T}
		\end{equation}
	}
	\textcolor{black}{Here, $Z_{\rm J} (\omega,\Phi)$ characterizes the effective impedance of the SQUID, while $Z_{\rm K}$ accounts for the kinetic impedance of the superconducting loop wire. Consequently, the state of matching is achieved when $Z_{\rm K}=Z_{\rm J}=0$, while the resistors possess identical values, i.e., $R_D= R_S$. In practical terms, this means that the equal resistors are interconnected via lossless and negligible inductance lines. In this scenario, $\tau(\omega)= 1$ over the whole frequency range, and power is fully transmitted at the quantum limit of thermal conductance $\Dot{Q}_{\rm \nu}= \frac{\pi k_{\rm B}^2}{12\hbar}(T_{\rm D}^2 - T_{\rm S}^2 )$ \cite{Timofeev2009}. However, it is plausible to consider that the SQUID impedance does not inherently lead to $Z_J= 0$, akin to the insights in \cite{MatthiasMeschkeandPekolaJukka2006}, thereby resulting in an unfavorable matching condition. Therefore, in our design,  $Z_{\rm K}(\omega)\approx 0$ for $\omega\lesssim k_B T/\hbar $. Hence, any deviation from the quantum limit of heat transfer can be attributed to the physics occurring at the SQUID.}